\documentclass[%
preprint,
 amsmath,amssymb,
 aps,
]{revtex4-1}

\usepackage{graphicx}
\usepackage{dcolumn}
\usepackage{bm}
\usepackage{xcolor}
\usepackage{qcircuit}
\usepackage{braket}
\graphicspath{ {./images/} }
\usepackage{subcaption}
\usepackage[font=small,labelfont=bf,
  justification=justified]{caption}
\usepackage{hyperref}

\renewcommand{\figureautorefname}{Figure~\negthinspace}

\begin{document}

\preprint{BNL}

\title{Quantum machine learning with differential privacy}

\author{William M Watkins}
\email{wwatkins@bnl.gov}
\affiliation{%
 Computational Science Initiative, Brookhaven National Laboratory, Upton, NY 11973, USA
}%
\author{Samuel Yen-Chi Chen}%
\email{ychen@bnl.gov}
\affiliation{%
 Computational Science Initiative, Brookhaven National Laboratory, Upton, NY 11973, USA
}%
\author{Shinjae Yoo}%
\email{sjyoo@bnl.gov}
\affiliation{%
 Computational Science Initiative, Brookhaven National Laboratory, Upton, NY 11973, USA
}%



\date{\today}

\begin{abstract}
Quantum machine learning (QML) can complement the growing trend of using learned models for a myriad of classification tasks, from image recognition to natural speech processing. A quantum advantage arises due to the intractability of quantum operations on a classical computer. Many datasets used in machine learning are crowd sourced or contain some private information. To the best of our knowledge, no current QML models are equipped with privacy-preserving features, which raises concerns as it is paramount that models do not expose sensitive information. Thus, privacy-preserving algorithms need to be implemented with QML. One solution is to make the machine learning algorithm differentially private, meaning the effect of a single data point on the training dataset is minimized. Differentially private machine learning models have been investigated, but differential privacy has yet to be studied in the context of QML. In this study, we develop a hybrid quantum-classical model that is trained to preserve privacy using differentially private optimization algorithm. This marks the first proof-of-principle demonstration of privacy-preserving QML. The experiments demonstrate that differentially private QML can protect user-sensitive information without diminishing model accuracy. Although the quantum model is simulated and tested on a classical computer, it demonstrates potential to be efficiently implemented on near-term quantum devices (noisy intermediate-scale quantum [NISQ]). The approach's success is illustrated via the classification of spatially classed two-dimensional datasets and a binary MNIST classification. This implementation of privacy-preserving QML will ensure confidentiality and accurate learning on NISQ technology.




\end{abstract}

\keywords{Quantum machine learning, noisy intermediate-scale quantum, variational quantum classifiers, differential privacy}
\maketitle


\section{\label{sec:Introduction}Introduction}

Recent advances in machine learning (ML), particularly deep learning (DL), have been successfully applied to computer vision \cite{simonyan2014very, Szegedy2014GoingConvolutions, Voulodimos2018DeepReview}, natural language processing \cite{Sutskever2014SequenceNetworks}, and even toward playing the game of \textit{Go} \cite{Silver2016MasteringSearch}. Notably, DL has been able to perform certain tasks with superhuman performance.
%

Concurrently, quantum computing machines have been introduced to the market by several tech companies. These machines are noisy and do not run in a fault-tolerant manner. Hence, they are referred to as \textit{noisy intermediate-scale quantum} (NISQ) devices \cite{Preskill_2018}. However, it has been shown that even these near-term machines can perform several calculations better than their classical counterparts \cite{arute2019quantum}. Various quantum algorithms have been developed to harness the power of these near-term quantum devices, including the \emph{variational algorithm}, which has been successful in calculating chemical ground states \cite{peruzzo2014variational}, optimization problems \cite{cerezo2020variational, bharti2021noisy} and certain ML tasks \cite{mitarai2018quantum, schuld2018circuit, Farhi2018ClassificationProcessors, benedetti2019parameterized, mari2019transfer, abohashima2020classification, easom2020towards, sarma2019quantum, chen2020hybrid, stein2020hybrid,chen2020quantum,kyriienko2020solving,dallaire2018quantum, stein2020qugan, zoufal2019quantum, situ2018quantum,nakaji2020quantum,lloyd2020quantum, nghiem2020unified,chen19, lockwood2020reinforcement, jerbi2019quantum,bausch2020recurrent,yang2020decentralizing}.
%

With advances in quantum computing capabilities, a growing number of ML tasks are expected to be implemented on quantum computers. Current successful ML models rely on massive datasets, and quantum machine learning (QML) is no exception. 
Most data used for building state-of-the-art ML models are collected from users. However, sensitive data, for example, personal video and voice recordings, medical records, and financial data, should never be accessible by unauthorized third-party users. Even if malicious adversaries cannot directly access the training data, they still may deduce a given data entry by attacking the trained model. One of the simplest privacy attacks is \emph{membership inference}, where the adversary attempts to predict if a given example was in the training set. In \cite{choquettechoo2021labelonly}, the authors demonstrate that membership inference attacks are robust against defense measures, such as confidence masking. Only differentially private training and high-level $\ell_{2}$ regularization can properly screen for (or safeguard against) such attacks.

Revealing private information is a significant problem for language models, such as GPT-2, as many are trained with either private text or sensitive public text \cite{carlini2019secret}. In \cite{carlini2020extracting}, it showed that training data can be extracted by carefully analyzing and sampling outputs, even for models exponentially smaller than the training set. This kind of \emph{data extraction attack} is not the only type that can result from ``black box'' access. \emph{A model-inversion attack} successfully recovered images from a facial recognition algorithm in \cite{fredriksonModelInversion} with only access to a person's name and the confidence levels outputted from the ``black box'' model. Furthermore, in many applications, a hostile adversary also may have access to the model parameters. In mobile applications, the model usually is stored on the device to reduce communication with a central server \cite{deepLearningDP}. \emph{Differential privacy} (DP) is an optimization framework to address these issues.

%
%
DP involves a trade-off of accuracy and power to protect the identity of data. Differentially private QML will allow private and efficient processing of big data. We hypothesize that the benefits of QML will offset the decrease in accuracy arising from DP.
This research aims to create a hybrid quantum-classical model based on a variational quantum circuit (VQC) and train it using a differentially private classical optimizer. 
%
The classification of two-dimensional (2D) data to two classes is used to test the efficiency of the DP-VQC. As controls in the experiment, we will compare its accuracy to classical neural networks (with and without DP) and a non-private quantum circuit. Two classification tasks are used as benchmarks to compare the efficiencies of private and non-private VQCs to their classical analogs.

The novel work detailed in Section~\ref{sec:DifferentialPrivacyQuantum} represents the main contribution of this research, exploring how we develop a novel framework that ensures privacy-preserving QML and employ it in two benchmark examples (as follows):
\begin{itemize}
    \item Demonstrate differentially private training on VQC-based ML models.
    \item Demonstrate that a ($\varepsilon$,$10^{-5}$)-DP VQC trains to accuracies exceeding 90\% for an MNIST task with $\varepsilon$ between 0.5 and 1.0.
\end{itemize}
%

Section~\ref{sec:Background} introduces the concept of differentially private ML and the required QML background. Section~\ref{sec:DifferentialPrivacyQuantum} illustrates the proposed differentially private QML.  Section~\ref{sec:ExpAndResults} describes the experimental settings and performance of the proposed differentially private quantum learning and is followed by additional discussions in Section~\ref{sec:Discussion}. Section~\ref{sec:Conclusion} is the conclusion.

\section{\label{sec:Background}Background}
%
%
\subsection{\label{sec:MachineLearning}Supervised Learning}

\emph{Supervised learning} is an ML paradigm that learns or trains a function that maps the input to output given the input-output pairs~\cite{russell2002artificial}. That is, given the training dataset $\{(\bm{x_i},\bm{y_i})\}$, it is expected that after successful training, the learned function $f_{\theta}$ is able to output the correct or approximate value $\bm{y_j}$ provided the testing case $\bm{x_j}$.
To make the training possible, we must specify the \emph{loss function} or \emph{cost function} $L(\hat{\bm{y}}, \bm{y})$, which defines how close the output of the ML model $\hat{\bm{y}} = f_{\theta}(\bm{x})$ is to the ground truth $\bm{y}$.
The \emph{learning} or \emph{training} of an ML model generally aims to minimize the loss function.

In classification tasks, the model is trained to output discrete labels or the targets $\bm{y}$ given the input data $\bm{x}$. 
For example, in computer vision applications, it is common to train ML models to classify images. The most famous example is the MNIST dataset~\cite{lecun1998mnist}. In MNIST, there are around $5000$ images of handwritten digits of the numbers $0$-$9$. In this case, the ML model is trained to output the probability distribution $P(y_i \vert \bm{x})$. Here, $P(y_i \vert \bm{x})$ represents the probability of label $y_i$ of each number $i \in \{0 \cdots 9 \}$ given the input data, which is a image in this scenario.

In classification, the \emph{cross-entropy loss} is the common choice for the loss function. It can be written in the following formulation:
\begin{equation}
    L(\hat{\bm{y}}, \bm{y}) = -\sum_{c=1}^{M} y_{o, c} \log \left(\hat{y}_{o, c}\right),
\end{equation}
where
\begin{itemize}
    \item $M$ = the number of classes.
    \item $log$ = the natural log.
    \item $y_{o, c}$ = the binary indicator ($0$ or $1$) if class label $c$ is the correct classification for observation $o$.
    \item $\hat{y}_{o, c}$ = the predicted probability observation $o$ is of class $c$.
\end{itemize}

The loss function then is used to optimize the model parameters $\theta$. In the current DL practice, the model parameters are updated via various gradient descent methods \cite{ruder2016overview}. The ``vanilla'' form of gradient descent is: 
\begin{equation}
    \theta \leftarrow \theta - \eta \nabla_{\theta} L(f_{\theta}(\bm{x}),\bm{y}),
\end{equation}
where $\theta$ is the model parameter, $L$ is the loss function, and $\eta$ is the learning rate or the step-size of each updating step.
Mini-batch stochastic gradient descent (SGD) simplifies ML by approximating the loss gradient when the dataset is large or when it is impractical to calculate the loss for the whole dataset at once. Suppose the training data include $N$ points, then define a randomly sampled subset of points $B$. This is the mini-batch. Equation \ref{eq:BatchGrad} approximates the gradient from the whole training set $\frac{1}{N} \sum_i \nabla_{\theta} L(f_{\theta}(\bm{x}_i),\bm{y}_i)$ with a loss gradient calculated for a subset of the training set, the mini-batch.
 \begin{equation} \label{eq:BatchGrad}
    \mathbf{g}_B = \sum_{i \in B} \frac{1}{|B|} \nabla_{\theta} L(f_{\theta}(\bm{x}_i),\bm{y}_i),
 \end{equation}
where $B$ is the mini-batch set randomly sampled from the complete set of inputs and associated ground truth labels. This batch gradient is used in the step update rule instead of the total loss gradient $\theta \leftarrow \theta - \eta\mathbf{g}_B$. The batch gradient is recalculated $N/|B|$ times per epoch, and the model parameters are updated for each gradient batch.

However, this vanilla form does not always work. For example, it may be easily stuck in local optima \cite{ruder2016overview}, or it can make the model difficult to train or converge. There are several gradient-descent variants that are successfully applied in DL \cite{ruder2016overview, Tieleman2012, kingma2014adam}. Based on previous works \cite{chen2020quantum, chen19}, we use the RMSProp optimizer to optimize our hybrid quantum-classical model.
RMSProp \cite{Tieleman2012} is a special kind of gradient-descent method with an adaptive learning rate that updates the parameters $\theta$ as:
\begin{subequations}
\begin{align}
        E\left[g^{2}\right]_{t} &= \alpha E\left[g^{2}\right]_{t-1}+ (1 - \alpha) g_{t}^{2}, \\ 
        \theta_{t+1} &= \theta_{t}-\frac{\eta}{\sqrt{E\left[g^{2}\right]_{t}}+\epsilon} g_{t},
\end{align}
\end{subequations} 
where $g_t$ is the gradient at step $t$ and $E\left[g^{2}\right]_{t}$ is the weighted moving average of the squared gradient with $E[g^2]_{t=0} = g_0^2$. In this paper, the hyperparameters are set for all experiments as follows: learning rate $\eta =0.05$, smoothing constant $\alpha = 0.9$, and $\epsilon = 10^{-8}$.


\subsection{\label{sec:QuantumComputingBasics}Quantum Computing Basics}
Because of the power of superposition and entanglement generated by quantum gates, quantum computing can create a huge speedup in certain difficult computational tasks and afford quantum advantages to ML \cite{nielsen2002quantum,Schuld_2019}. 
%
A \emph{qubit} is the basic unit of quantum information processing that can consist of any two state system, i.e., the spin of an electron or polarization of a photon. Such a state will be written as $\ket{\psi}=\alpha\ket{1}+\beta\ket{0}$, where the probability of measuring $\ket{1}$ and $\ket{0}$ is $|\alpha|^2$ and $|\beta|^2$, respectively.

Because all classical operations can be considered a set of reversible logical operations, analogous quantum operations can be formalized. These operators are unitary and can be thought of as successive rotations, such that the logic operators are equivalent to quantum rotations. 
The basic components of quantum rotations are the Pauli matrix,
\begin{equation}
    \mathbb{I} = 
   \begin{bmatrix}
    1 & 0 \\
    0 & 1 
    \end{bmatrix},
    \sigma_x = 
   \begin{bmatrix}
    0 & 1 \\
    1 & 0 
    \end{bmatrix},
    \sigma_y = 
   \begin{bmatrix}
    0 & -i \\
    i & 0 
    \end{bmatrix},
    \sigma_z = 
   \begin{bmatrix}
    1 & 0 \\
    0 & -1 
    \end{bmatrix}.
\end{equation}
With the Pauli matrix, we can define the single-qubit rotation along each of the $X$, $Y$, and $Y$-axis as follows:
\begin{equation}
\begin{split}
R_x(\phi) &= e^{-i\phi\sigma_x/2} = \begin{bmatrix}
    \cos(\phi/2) & -i\sin(\phi/2) \\
    -i\sin(\phi/2) & \cos(\phi/2)
    \end{bmatrix} \\
R_y(\phi) &= e^{-i\phi\sigma_y/2} = \begin{bmatrix}
    \cos(\phi/2) & -\sin(\phi/2) \\
    \sin(\phi/2) & \cos(\phi/2)
\end{bmatrix} \\
R_z(\phi) &= e^{-i\phi\sigma_z/2} = \begin{bmatrix}
    e^{-i\phi/2} & 0 \\
    0 & e^{i\phi/2}
\end{bmatrix}.
\end{split}
\end{equation}
The general single-qubit rotation can be constructed with two of the single-qubit rotations $R_{x}$, $R_{y}$, and $R_{z}$.
\begin{equation}
\label{eqn:generalRotOperation}
    R(\phi,\theta,\omega) = R_{z}(\omega)R_{y}(\theta)R_{z}(\phi)= \begin{bmatrix}
e^{-i(\phi+\omega)/2}\cos(\theta/2) & -e^{i(\phi-\omega)/2}\sin(\theta/2) \\
e^{-i(\phi-\omega)/2}\sin(\theta/2) & e^{i(\phi+\omega)/2}\cos(\theta/2)
\end{bmatrix}.
\end{equation}
For example, the quantum NOT gate also is known as the ``Pauli-$X$ gate,'' which corresponds to a $\pi$ rotation about the $X$-axis \cite{Steane_1998}.
\begin{subequations}
\begin{equation}
    U_{NOT}\ket{1} = \ket{0}; U_{NOT}\ket{0} = \ket{1}
\end{equation}
\begin{equation}
    U_{NOT}= e^{-i\pi\sigma_x} =
    \begin{pmatrix}
    0 & 1 \\
    1 & 0 
    \end{pmatrix}.
\end{equation}  
\end{subequations}
 The true power of quantum computing stems from quantum entanglement, which can be achieved by using two-qubit quantum gates. The controlled-NOT (CNOT) gate, shown in Eq. \ref{eqn:CNOTgate}, is a gate commonly used to entangle qubits. It reverses the state of second qubit if the first qubit (\emph{control qubit}) is in the $\ket{1}$ state. 
\begin{equation}
    \label{eqn:CNOTgate}
    U_{CNOT} = \begin{bmatrix}
                    1 & 0 & 0 & 0 \\
                    0 & 1 & 0 & 0 \\
                    0 & 0 & 0 & 1 \\
                    0 & 0 & 1 & 0 
                \end{bmatrix}.
\end{equation}
%
Its operation on the quantum state can be described in the following circuit diagram: \\
\[
\Qcircuit @C=1em @R=1em {
\lstick{\ket{\Psi}} & \ctrl{1} & \qw \\
\lstick{\ket{0}}    & \targ    & \qw
}
\]
where $\ket{\Psi}$ is a single-qubit state.
Concretely, if the $\ket{\Psi}$ is in the state $\alpha \ket{0}+\beta \ket{1}$, which means the system is in $\ket{\Psi} \otimes \ket{0}$, then under the CNOT operation, the state will
\begin{equation}
\label{eqn:CNOTentanglement}
    \begin{split}
        U_{CNOT} \ket{\Psi}\otimes \ket{0}
        &= U_{CNOT}\left[\left( \alpha \ket{0} + \beta \ket{1}\right) \otimes \ket{0}\right] \\
        &= U_{CNOT} \left[ \alpha \ket{0} \otimes \ket{0} + \beta \ket{1} \otimes \ket{0}\right] \\
        &= \alpha U_{CNOT} \ket{0} \otimes \ket{0} + \beta U_{CNOT} \ket{1} \otimes \ket{0} \\
        &= \alpha  \ket{0} \otimes \ket{0} + \beta \ket{1} \otimes \ket{1}
    \end{split}.
\end{equation}
The set of CNOT and single-qubit rotation operators allows for a rich group of quantum algorithms that already have been shown to be faster than their classical counterparts, for example, in factorization problems \cite{Shor_1997} and database searching \cite{grover_1996}. 
The quantum algorithm output is the observation of the final quantum state. On a real quantum computing device, the expectation values can be retrieved through repeated measurements (\emph{shots}). In simulation, the expectation values $\bra{0} U_0^{\dagger}U_1^{\dagger} \cdots U_n^{\dagger}U_n \cdots U_0U_1 \ket{0}$ can be calculated analytically.
For a more detailed review of quantum computing, measurements, and algorithms, refer to \cite{Steane_1998, hey_1999, Ladd_2010}.


\subsection{\label{sec:VQCs}Variational Quantum Circuits}
In recent years, quantum computing has become feasible due to many breakthroughs in condensed matter physics and engineering. Companies, such as IBM \cite{cross2018ibm}, Google \cite{arute2019quantum}, and D-wave \cite{grzesiak2020efficient}, are creating NISQ devices \cite{Preskill_2018}. However, noise limits the reliability and scalability in which quantum circuits can be used. For example, quantum algorithms requiring large numbers of qubits or circuit depth cannot be faithfully implemented on these NISQ devices. Because current cloud-based quantum devices are not suitable for the training described in this research, quantum circuit simulators are used \cite{schuld2019evaluating}. 

\begin{figure}[htbp]
\begin{center}
\begin{minipage}{10cm}
\Qcircuit @C=1em @R=1em {
\lstick{\ket{0}} & \gate{R_y(\arctan(x_1))} & \gate{R_z(\arctan(x_1^2))} & \ctrl{1}   & \qw        & \targ     & \gate{R(\alpha_1, \beta_1, \gamma_1)} & \meter \qw \\
\lstick{\ket{0}} & \gate{R_y(\arctan(x_2))} & \gate{R_z(\arctan(x_2^2))} & \targ      & \ctrl{1}  & \qw     & \gate{R(\alpha_2, \beta_2, \gamma_2)} &  \qw \\
\lstick{\ket{0}} & \gate{R_y(\arctan(x_3))} & \gate{R_z(\arctan(x_3^2))} & \qw        & \targ     & \ctrl{-2}    & \gate{R(\alpha_3, \beta_3, \gamma_3)} &  \qw 
\gategroup{1}{4}{3}{7}{.7em}{--}\qw 
}
\end{minipage}
\end{center}
\caption[Variational quantum circuit component.]{{\bfseries Variational quantum circuit component.}
  The single-qubit gates $R_y(\arctan(x_i))$ and $R_z(\arctan(x_i^2))$ represent rotations along the $y$- and $z$-axis by the given angle $\arctan(x_i)$ and $\arctan(x_i^2)$, respectively. Arctan is used because the input values are not in the interval of $[-1, 1]$. The CNOT gates are used to entangle quantum states from each qubit and $R(\alpha,\beta,\gamma)$ represents the general single qubit unitary gate with three parameters.
  The parameters labeled $R_y(\arctan(x_i))$ and $R_y(\arctan(x_i^2))$ are for state preparation and are not subject to iterative optimization. Parameters labeled $\alpha_i$, $\beta_i$ and $\gamma_i$ are optimized iteratively. The dashed box denotes one layer of a quantum subcircuit. The dial to the far right represents that the circuit has one output that is the $\sigma_z$ measurement of the first qubit.}

\label{Fig:Basic_VQC}
\end{figure}
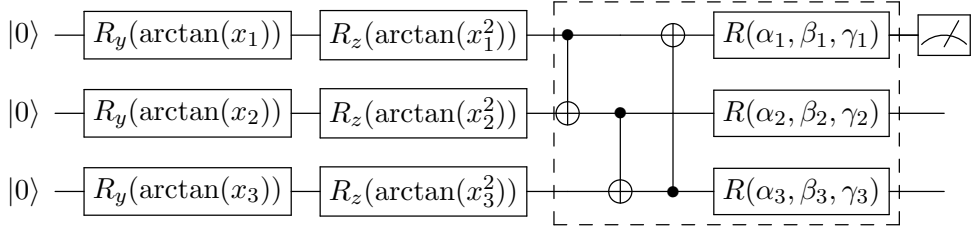

VQCs are a special kind of quantum circuit, equipped with \emph{tunable} or \emph{learnable} parameters that are subject to iterative optimization \cite{mitarai2018quantum, schuld2018circuit}. \figureautorefname{\ref{Fig:Basic_VQC}} presents the basic components of a VQC. 
VQCs potentially can be robust against device noise as they can absorb the noise effects into their parameters in the optimization process \cite{cerezo2020variational, bharti2021noisy}. Numerous efforts have been made to design quantum algorithms based on VQCs \cite{cerezo2020variational, bharti2021noisy}, including the calculation of chemical ground states \cite{peruzzo2014variational} and optimization problems \cite{farhi2014quantum}.


Several theoretical studies have shown that VQCs are more capable than conventional deep neural networks~\cite{sim2019expressibility,lanting2014entanglement,du2018expressive, abbas2020power} in terms of the number of parameters or convergence speed. Recent results have numerically demonstrated that certain quantum architectures can perform better than their classical counterparts under specific conditions. For example, quantum convolutional neural networks (QCNNs) can learn faster (with fewer training epochs) than classical CNNs and reach higher accuracies, even when the number of parameters are similar \cite{chen2020qcnn, chen2021qgcnn}. In \cite{chen2020quantum}, a demonstration shows that a quantum long short-term memory (LSTM) can learn much faster (i.e., reach comparable accuracies with fewer training epochs) than a classical LSTM in function approximation tasks when the number of parameters are similar.
%
VQCs have been applied in several classic ML tasks, such as classification \cite{mitarai2018quantum, schuld2018circuit, Farhi2018ClassificationProcessors, benedetti2019parameterized, mari2019transfer, abohashima2020classification, easom2020towards, sarma2019quantum, liu2019hybrid, stein2020hybrid,chen2020hybrid, chen2020qcnn, chen2021qgcnn}, function approximation \cite{chen2020quantum, mitarai2018quantum}, solving differential equations \cite{kyriienko2020solving}, sequential learning \cite{chen2020quantum, bausch2020recurrent, takaki2020learning}, and generative modeling \cite{dallaire2018quantum, stein2020qugan, zoufal2019quantum, situ2018quantum}. Recent results have demonstrated the successful application of VQCs in the forefront of ML, for example, in metric learning \cite{lloyd2020quantum, nghiem2020unified}, deep reinforcement learning \cite{chen19, lockwood2020reinforcement, jerbi2019quantum, wu2020quantum}, and speech recognition \cite{yang2020decentralizing}.

\subsection{\label{sec:DifferentialPrivacy}Differential Privacy}
%
\begin{figure}[htbp]
\includegraphics[width=0.5\linewidth]{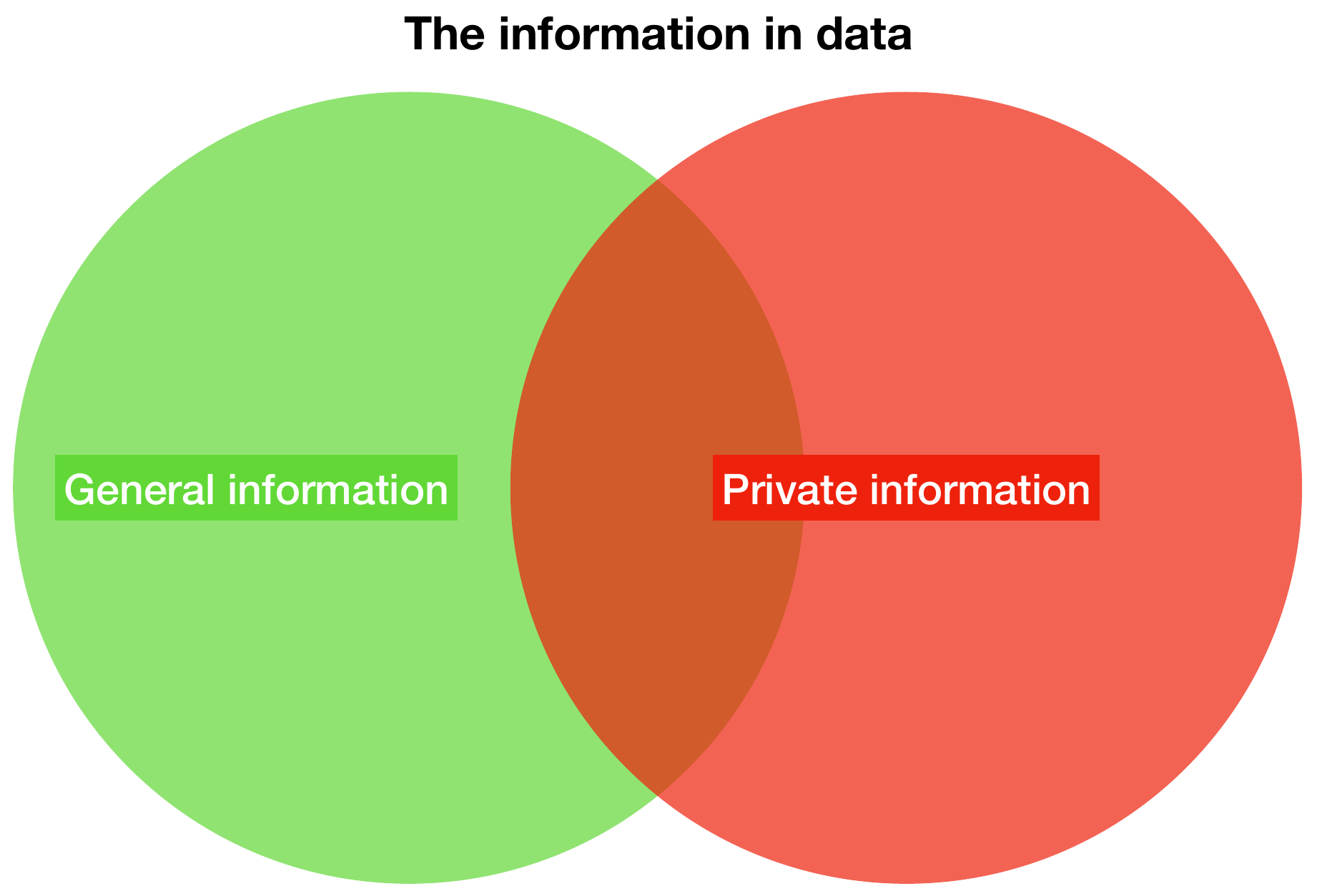}
\caption{{\bfseries Information in data under the view of differential privacy.} In a DP context, general information is that of the entire population in the data. On the other hand, private information is specific to a particular data entry.}
\label{fig:private_vs_general_info}
\end{figure}
\begin{figure}[htbp]
\includegraphics[width=0.8\linewidth]{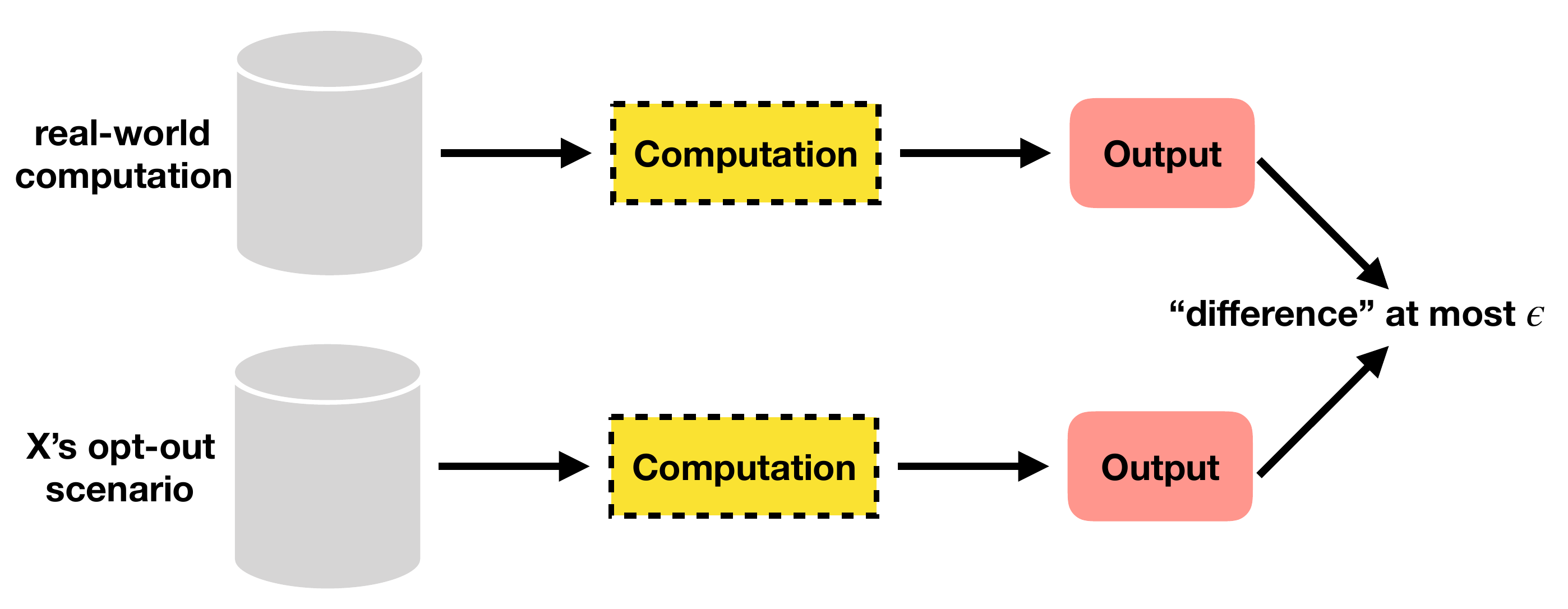}
\caption{{\bfseries Differential Privacy.} }
\label{fig:dp_concept}
\end{figure}
%

%
Many technology companies collect data about the online presence of their users, and these data are shared, sometimes publicly, to use in focused marketing. This can create a breach in privacy because anonymizing data requires more than just erasing the name from each data entry \cite{foundationDP}. Privacy also can be breached by ML models that use crowd-sourced information and data scraped from the Internet. Previous studies have shown that models memorize their training samples, and even models with millions of parameters can be attacked to output memorized data \cite{carlini2020extracting}.

Section~\ref{sec:Introduction} detailed the necessity of protecting information through privacy-preserving training algorithms. In other words, anonymizing data requires more than just censoring personally identifiable information (PII) from each data entry \cite{foundationDP}. The solution requires using DP to curtail privacy leaks. 

DP is a powerful framework to restrict the information that adversaries can obtain from attacking a trained ML model, but it is not an all-powerful technique. 
There are two kinds of information under the perspective of DP: \emph{general information} and \emph{private information}. General information refers to the information that does not specify any particular data entry and can be seen as the general property of the underlying population. On the other hand, private information refers to the information that is specific to any individual data entry (\figureautorefname{\ref{fig:private_vs_general_info}}).
For a concrete example \cite{foundationDP}, consider a study about smokers. An adversary may still learn information from the trained model, e.g., a differentially private query could show that smoking correlates to lung cancer, yet it is impossible to deduce whether or not a specific person is involved in the study. 
This is known as \emph{general information}. It remains possible to deduce that an individual smoker is likely to have lung cancer, but this deduction is not due to her/his presence in the study. DP does protect an individual's \emph{private information}. 
The power of DP is that deductions about an individual cannot be influenced by the fact that the person did or did not participate in the study \cite{foundationDP}. 

DP seeks to create a randomized machine, characterized by the hyperparamemers $\varepsilon$ and $\delta$, which gives roughly the same output for two similar datasets. In the context of ML, the output here is the \emph{trained model}. This means an adversary cannot deduce the dataset from the output even with auxiliary information or infinite computing resources. \figureautorefname{\ref{fig:dp_concept}} illustrates the concept of DP by comparing the output between two datasets, where one $X$ opts out of the dataset. Changing the input means the output could be very different, but DP ensures that the outputs only differ by, at most, $\varepsilon$. In other words, DP combats extraction attacks by having the output be just as likely produced from a model with or without a given training point \cite{DPReview}.

In DP, we are interested in mechanisms $\mathcal{M}$, which are randomized algorithms. Suppose $\mathcal{M}$ has a domain $A$ and a discrete range $B$. A randomized algorithm maps its domain $A$ to the probability space of $B$. Given an input $a \in A$, the algorithm $\mathcal{M}$ outputs $M(a) = b$ with probability $(M(a))_b$ for each $b \in B$. In general, a point in the domain may be a database (i.e., collection of records). A collection of records can be represented by a histogram, so the domain is the set of all possible histograms $\mathbb{N^{|\mathcal{X}|}}$. A $x \in \mathbb{N^{|\mathcal{X}|}}$ has $|\mathcal{X}|$ elements, where $x_i$ is the number of elements of type $i \in \mathcal{X}$. Additionally, there is a $\ell_{1}$-norm defined, such that $\|x-y\|_1 \leq 1$ represents the fact that $x$ and $y$ are neighboring databases, i.e., they differ by up to one record \cite{foundationDP}.

Rigorously, the definition of DP is \cite{foundationDP} a randomized algorithm $\mathcal{M}$ with a domain $\mathbb{N}^{|\mathcal{X}|}$ is $(\varepsilon, \delta)$-differentially private for all $S \subseteq \mathrm{range}(\mathcal{M})$ and for all $x, y \in \mathbb{N}^{|\mathcal{X}|}$, such that $\|x-y\|_1 \leq 1$: 
\begin{equation}
    Pr[\mathcal{M}(x) \in S] \leq \mathrm{exp}(\varepsilon) Pr[\mathcal{M}(y) \in S] + \delta,
\end{equation}
where
\begin{itemize}
    \item $\mathcal{M}$ = the randomized algorithm.
    \item $\mathbb{N^{|\mathcal{X}|}}$ = the set of records and the union of the input and label sets in ML context.
    \item $S$ = output randomized algorithm; some subset of all possible model configurations or parameters in ML context.
    \item $x$ = set of records used for model training.
    \item $y$ = another set of records for model training, neighboring $x$.
    \item $\varepsilon$ = privacy loss for the randomized algorithm.
    \item $\delta$ = cutoff on DP, the percentage chance that the model does not preserve privacy.
    
\end{itemize}
$(\varepsilon, \delta)$-DP is a relaxation of $\varepsilon$-DP because there is a chance $\delta$ that the privacy is broken. DP gives the worse-case scenario privacy loss, so a smaller epsilon does not necessarily mean the privacy is better. However, the additional noise typically means that accuracy is worse.

An important characteristic in determining the effectiveness of a differentially private algorithm is the \emph{privacy loss}. Privacy loss is defined for a given observation $\xi \in \mathrm{range}(\mathcal{M})$, which quantifies the likeness of observing $\xi$ from $\mathcal{M}(x)$ versus $\mathcal{M}(y)$ \cite{DPReview}.
\begin{equation}
    \mathcal{L}^{(\xi)}_{\mathcal{M}(x) || \mathcal{M}(y)} = \ln\left(\frac{Pr[\mathcal{M}(x)=\xi]}{Pr[\mathcal{M}(y)=\xi]}\right).
\end{equation}
Combining these two equations shows that a ($\varepsilon, \delta)$-differentially private algorithm has a privacy budget of $\varepsilon$. 

\subsection{\label{sec:DifferentialPrivacyMachineLearning}Differential Privacy in Machine Learning}
For ML, we can interpret the randomized algorithm $\mathcal{M}: A \to B$ as a training algorithm with a training set $x \in A$, which produces a model $b \in B$ \cite{deepLearningDP, disparateImpact}. The definition of DP implies that two training sets, which only differ by the omission of a record, should be equally likely to output a given model, i.e., the set of parameters completely describing the model.

The most basic technique to ensure DP is the \emph{Gaussian Mechanism} as defined in \cite{deepLearningDP, foundationDP, mcmahanGeneralApproachDPTraining}. Every deterministic function $f(d)$ has a defined sensitivity $S_f = \mathrm{max}(|f(d)-f(d')|)$ given that $d, d'$ are adjacent databases. Then, the Gaussian algorithm is $(\varepsilon, \delta)$-differentially private for some noise multiplier $\sigma$, such that:
\begin{equation} \label{eq:GaussMech}
    \begin{aligned}
    \mathcal{M}(d) = f(d) + \mathcal{N}(0, S_f^2 \sigma^2 \mathbb{I}), \\
    \delta \geq \frac{4}{5}e^{-{(\sigma \varepsilon)}^2/2.}
    \end{aligned}
\end{equation}
There is an infinite number of pairs $(\varepsilon, \delta)$, which can be defined for a given noise multiplier $\sigma$, although usually, as in \cite{disparateImpact}, $\delta$ will be defined as a constant.
Likewise, for ML, the most important techniques for creating DP are to add Gaussian noise, as well as to clip the loss gradients \cite{deepLearningDP}. The gradient clip reduces the effect any single data entry can have on the model training, making membership inference difficult. The hyperparameters associated with these operations are the noise multiplier, $\sigma$, and a cutoff for the $\ell_{2}$ norm, $S$ \cite{disparateImpact, deepLearningDP}. After calculating the gradients, if the batch gradient has an $\ell_{2}$ norm greater than the cutoff, it is scaled down to have a norm equal to the cutoff. After clipping the gradient, the gradient for the mini-batch has Gaussian noise added with a standard deviation equal to the $\ell_{2}$ norm cutoff multiplied by the noise factor, $\sigma$.
\begin{equation}
    \mathbf{g}_B \leftarrow \left[ \mathbf{g}_B*\mathrm{min}\left(1, \frac{S}{||\mathbf{g}_B||}\right) + \mathcal{N}(0,\sigma^2 S^2 \mathbb{I} ) \right].
\end{equation}
As in Equation \ref{eq:GaussMech}, a relationship between $\varepsilon, \delta, \sigma, and S$ exists, but its calculation is beyond the scope of this review. More information about the privacy loss calculator can be found in \ref{app:PyvacyLossCalc}. This modification to the optimizer algorithm can be applied to any ML algorithm (SGD, Adam, RMSprop, etc.). The DP-SGD algorithm is based on the techniques from \cite{mcmahanGeneralApproachDPTraining, deepLearningDP}. Details specific to the software package used in this study to implement DP, \texttt{PyVacy}, are available in \ref{app:PyvacyML}.

\begin{figure}[t]
\includegraphics[width=0.5\linewidth]{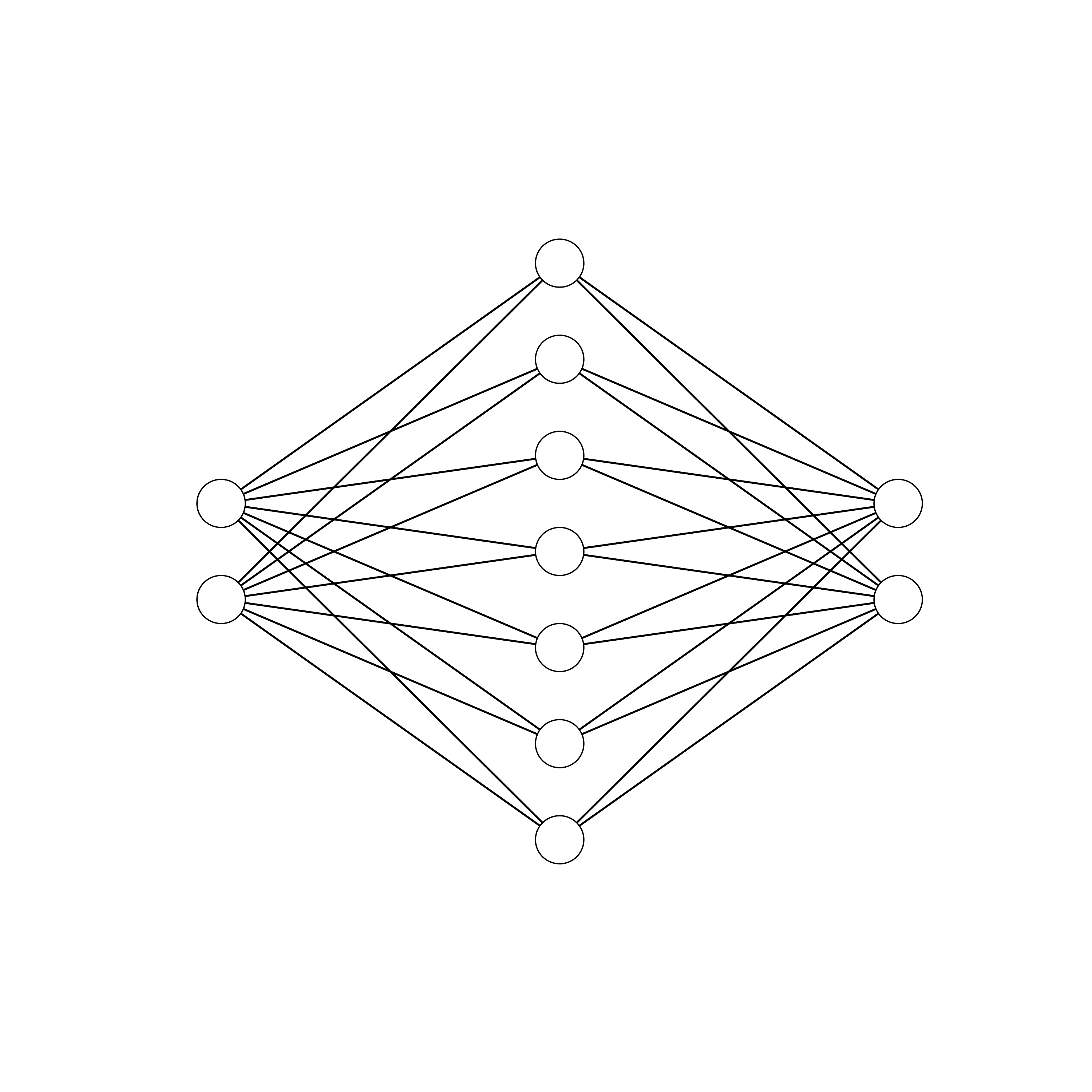}
\caption{Architecture for the classical neural network used as a control. The left layer is the 2D input, while the right layer is the output. The output is a vector of the probability of being in each class given the input.}
\label{fig:nn_arch}
\end{figure}

\section{\label{sec:DifferentialPrivacyQuantum}Differential Privacy in Quantum Classification}
\begin{figure}[htbp]
\includegraphics[width=0.8\linewidth]{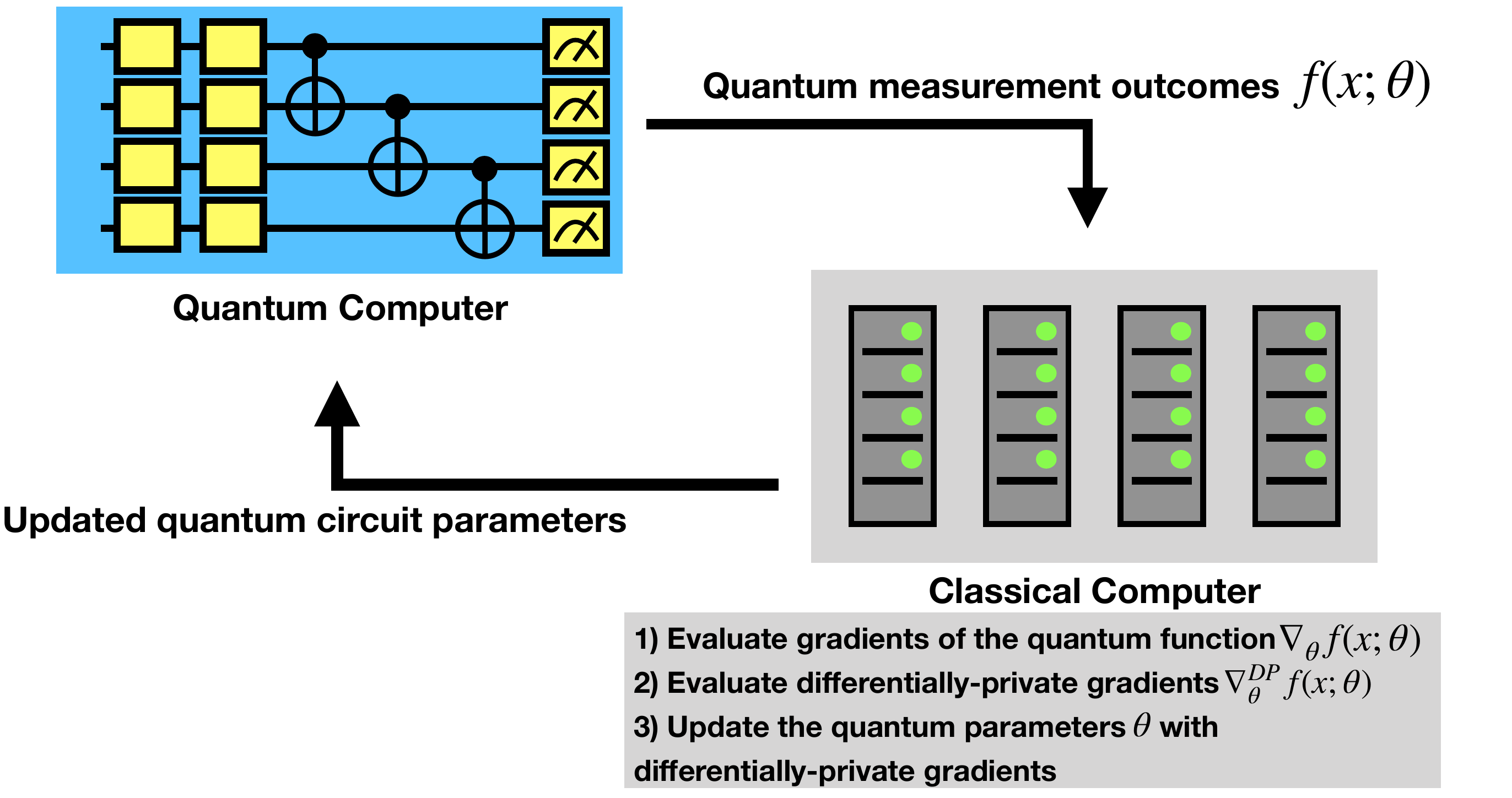}
\caption{{\bfseries Differential Privacy in Quantum Machine Learning.} In the proposed framework, the outputs from the quantum circuit are processed on a classical computer. The gradients of the quantum function $\nabla_{\theta}f(x;\theta)$ and the differentially private gradients $\nabla_{\theta}^{DP}f(x;\theta)$ are calculated. The quantum circuit parameters are updated according to the differentially private gradients and fed back to the quantum computer.}
\label{fig:dp_qml_concept}
\end{figure}

%
%
In this work, we propose a hybrid quantum-classical framework interfacing the differentially private classical optimization algorithms with VQC-based QML algorithms. In a hybrid quantum-classical model architecture, the quantum circuits are used to generate the output, mostly in the form of quantum measurement. The measured expectation values then can be used to evaluate the \emph{loss function} on a classical computer, which then will be used to evaluate the model's performance and adjust the circuit parameters. The updated circuit parameters are then fed back to the quantum computer. This iterative process gradually \emph{trains} the quantum circuit to achieve the desired results. The DP training in such a hybrid quantum architecture exists in the gradient calculation process, which is on the classical computer. \figureautorefname{\ref{fig:dp_qml_concept}} presents the proposed scheme.
%
%
\subsection{Quantum Encoding}
A quantum circuit operates on the quantum state. To make QML useful, the first step is to encode the classical data into a quantum state. 
\subsubsection{\label{sec:AmplitudeEncoding}Amplitude Encoding}
\emph{Amplitude encoding} is a technique to encode the classical vector $(\alpha_{0} \cdots \alpha_{2^n-1})$ into an $n$-qubit quantum state $\ket{\Psi} = \alpha_{0}\ket{00\cdots 0} + \cdots + \alpha_{2^n-1}\ket{11\cdots 1}$. The advantage of using this encoding method is that it is possible to significantly reduce the number of qubits and potentially the number of parameters of the quantum circuit. An $N$-dimensional input vector would require only $\log_{2}N$ qubits to encode. Refer to \cite{Schuld2018InformationEncoding, mottonen2005transformation} for details regarding this encoding procedure.
\subsubsection{\label{sec:VariationalEncoding}Variational Encoding}
In \emph{variational encoding}, the input values are used as the quantum rotation angles. A single-qubit gate with rotation along the $j$-axis by angle $\alpha$ is given by:
\begin{equation}
  \label{eq:SingleQubitRotation}
  R_j(\alpha)=e^{-i\alpha\sigma_j/2}=\cos\frac{\alpha}{2} I-i\sin\frac{\alpha}{2}\sigma_j,
\end{equation}
where $I$ is the identity matrix and $\sigma_{j}$ is the Pauli matrix with $j = x, y, z$. In this work, given a vector input $x_N$ with $N$ dimensions, we rotate each qubit by $R_i(x), i \in [0,N)$:
\begin{equation}
  \label{eq:StudyQubitRotation}
    \begin{aligned}
    R_i(x)=e^{-i\beta_i\sigma_z/2}e^{-i\alpha_i\sigma_y/2}, \\
     \alpha_i = \arctan(x_i);  \beta_i = \arctan(x_i^2).
     \end{aligned}
\end{equation}
Each single-qubit state is initialized by rotations in the $y$-axis then in the $z$-axis. This allows our inputs, $x \in X$, to be encoded into a quantum state of $N$ qubits. \figureautorefname{\ref{Fig:Basic_VQC}} depicts this particular encoding scheme. For a detailed review of different quantum encoding schemes, refer to \cite{Schuld2018InformationEncoding}.
\subsection{Quantum Gradients}
Modern DL practices heavily depend on gradient-based optimization methods. Classically, the gradients of DL models are calculated using \emph{backpropagation} methods \cite{lecun1989backpropagation}. In QML, the corresponding method is the \emph{parameter-shift rule}, which can calculate the analytical gradients of quantum models \cite{schuld2019evaluating, mitarai2018quantum}. 

For parameter-shift rule, knowledge of certain observables are given. A VQC's output can be modeled as a function of its parameters $f(x; \theta)$ with parameters $\theta$. Then, in most cases, the partial derivative of the VQC, $\nabla_{\theta} f(x; \theta)$, can be evaluated with the same quantum circuit only with the parameters shifted \cite{mitarai2018quantum}. We illustrate the procedure as follows: consider a quantum circuit with a parameter $\theta$, and the output can be modeled as the expectation of some observable, e.g., $B$ for some prepared state $\ket{\psi}=U(\theta)U_0(x)\ket{0}$ or $f(x; \theta) = \bra{0}U^{\dagger}_0(x)U^{\dagger}(\theta)\hat{B}U(\theta)U_0(x)\ket{0}$. This is simplified by considering the first unitary operation as preparing the state $\ket{x}$ and the other unitary operators as a linear transformation of the observable, $U^{\dagger}(\theta)\hat{B}U(\theta) = \mathcal{M}_{\theta} (\hat{B})$.
\begin{subequations}
\begin{equation}
    \label{eqn:quantumFunc}
    \begin{aligned}
    f(x; \theta) = \bra{x}\mathcal{M}_{\theta} (\hat{B})\ket{x}, \\
    \nabla_\theta f(x; \theta) = \bra{x} \nabla_\theta \mathcal{M}_{\theta} (\hat{B})\ket{x},
    \end{aligned}
\end{equation}
\begin{equation}
    \label{eqn:paramShift}
    \nabla_\theta \mathcal{M}_{\theta} (\hat{B}) = c[\mathcal{M}_{\theta+s} (\hat{B}) - \mathcal{M}_{\theta-s} (\hat{B})].
\end{equation}
\end{subequations}
It can be shown that a finite parameter, $s$, exists, such that the Equation \ref{eqn:paramShift} stands \cite{mitarai2018quantum}. This implies that the quantum circuit can be shifted to allow for a calculation of the quantum gradient with the same circuit.

%
%
Now that DP and our VQC architecture are introduced, we unveil our differentially private optimization algorithm---the first of its kind to ensure privacy-preserving QML. Our differentially private optimization framework starts by calculating the quantum gradient using the parameter shift rule. Next, we apply Gaussian noise and clipping mechanisms to this gradient, $\nabla_{\theta} f(x; \theta)$. The differentially private gradient, $\nabla_{\theta}^{DP} f(x; \theta)$, now is used in the parameter update step instead of the non-private gradient. This parameter update rule can be SGD, adaptive momentum, or RMSprop. In this study, we solely use RMSprop to update parameters.

\begin{equation}
    \label{eqn:quantumGradDP}
    \nabla_{\theta}^{DP} f(x; \theta) = \left[ \bra{x} \nabla_{\theta} \mathcal{M}_{\theta} (\hat{B})\ket{x}*\mathrm{min}\left(1, \frac{S}{|\bra{x} \nabla_{\theta} \mathcal{M}_{\theta} (\hat{B})\ket{x}|}\right) + \mathcal{N}(0,\sigma^2 S^2 \mathbb{I} ) \right],
\end{equation}    
where $\mathcal{M}_{\theta} (\hat{B})$ is defined in Equation \ref{eqn:quantumFunc} and $S, \sigma$ are the hyperparameters implicitly defining the level of privacy $(\varepsilon, \delta)$. This novel framework seamlessly incorporates privacy-preserving algorithms into the training of a VQC, ensuring $(\varepsilon, \delta)$-differential privacy. In this work, we choose the standard classification task to demonstrate the proof-of-concept result. However, the proposed framework is rather generic and can be applied to any hybrid quantum-classical ML scenarios.  
\section{\label{sec:ExpAndResults}Experiments and Results}
To demonstrate the hypothesized quantum advantage, this study compares differentially private VQCs (DP-VQCs) to non-private VQCs, as well as private and non-private neural networks. We also illustrate the efficacy of our differentially private QML framework. 
%
Two different types of classifications will be investigated as benchmarks: 1) labeling points in a 2D plane and 2) a binary classification from an MNIST dataset, differentiating between the `0' and `1' digits. The 2D datasets are standard benchmarks from \texttt{scikit-learn} \cite{scikit-learn} that are useful in QML because the inputs are low dimensional, thus easy to simulate on classical computers \cite{Schuld_2019}. Meanwhile, the MNIST dataset is used to study the performance of the proposed model with larger dimensional inputs.

We implement the model with several open-source software packages. The high-level quantum algorithms are implemented with \texttt{PennyLane}~\cite{bergholm2018pennylane}. The quantum simulation backend is \texttt{Qualacs} \cite{suzuki2020qulacs}, which is a high-performance choice when the number of qubits is large. The hybrid quantum-classical model is built with the \texttt{PyTorch} interface \cite{NEURIPS2019_9015}. For differentially private optimization, we employ the \texttt{PyVacy} package \cite{pyvacy}. 

The experiments are characterized by the hyperparameters of the neural network training process: the optimizer, number of epochs, number of training samples, learning rate, batch size, momentum, and weight penalty. When differentially private optimizers are used, the additional hyperparameters needed are the $\ell_{2}$ norm clip, noise multiplier, number of iterations, and $\varepsilon$. After preliminary experiments, the RMSprop optimizer was selected for use in all of the experiments presented in this paper. Most of the model's hyperparameters are the same for both the MNIST and scikit 2D set classification tasks. The learning rate is set to $0.05$, while the portion of training and testing is 60\% and 40\%, respectively. In addition, the batch size used is $32$ with a momentum value of $0.5$, but no weight regularization is used. 

An $\varepsilon$ is calculated from the DP hyperparameters, $S, \sigma$. Because all tasks are classifications, cross-entropy is used as the loss function for all training. According to \cite{DPReview, foundationDP}, the probability of breaking $\varepsilon$-DP should be $\delta \sim \mathcal{O}(1/n)$ for $n$ samples. A $\delta$ larger than $1/n$ always will be able to satisfy DP simply by releasing $n\delta$ complete records. Therefore, $\varepsilon$ is determined by hyperparameter choice, and $\delta$ is set to be $10^{-5}$ for the entire study.

\begin{table}[ht]
\begin{center}
 \begin{tabular}{||c || c | c | c || c | c | c | c | c | c ||} 
 \hline
 Exp & LR & Mom. & Batch & $\ell_{2}$ Clip & Noise Mult. & \# of Iter. & $\delta$ \\ [0.5ex] 
 \hline\hline
 non-private & 0.05 & 0.5 & 32 & n/a & n/a & n/a & n/a \\
 \hline
 DP & 0.05 & 0.5 & 32 & 1.0 & varies & 5 & $10^{-5}$ \\
 \hline
\end{tabular}
\end{center}
\caption{Hyperparameters chosen for non-private and differentially private classifiers. The neural networks and VQCs use the same hyperparameters for both classification tasks. The learning rate (LR) is the same across all experiments. Different noise multipliers are used to compare differentially private networks. The ``varies'' noise parameter means that multiple values of noise have been used in the DP-neural network and DP-VQC experiments. $\varepsilon$ also varies among DP experiments as it directly depends on the noise multiplier. The last four hyperparameters are applicable only with differentially private optimizers.}
\label{table:hyperparameters}
\end{table}

As part of the investigation into differentially private QML, classical and quantum classifiers are compared. For both the MNIST and 2D classifiers, the quantum circuit has two modules that contain the parameters for the unitary transforms comprising the two quantum subcircuits. 
\subsection{Two-dimensional Mini-benchmark Datasets}
\begin{figure}[htbp]
\begin{center}
\begin{minipage}{10cm}
\Qcircuit @C=1em @R=1em {
\lstick{\ket{0}} & \gate{R_y(\arctan(x_1))} & \gate{R_z(\arctan(x_1^2))} & \ctrl{1}       & \targ  & \qw       & \gate{R(\alpha_{i1}, \beta_{i1}, \gamma_{i1})} & \meter \qw \\
\lstick{\ket{0}} & \gate{R_y(\arctan(x_2))} & \gate{R_z(\arctan(x_2^2))} & \targ      & \ctrl{-1}  & \qw     & \gate{R(\alpha_{i2}, \beta_{i2}, \gamma_{i2})} & \meter \qw 
\gategroup{1}{4}{2}{7}{.7em}{--}\qw 
}
\end{minipage}
\end{center}
\caption[First VQC block for scikit 2D data classification]{{\bfseries First quantum circuit block for 2D classification.}
  The single-qubit gates $R_y(\arctan(x_i))$ and $R_z(\arctan(x_i^2))$ represent rotations along the $y$-axis and $z$-axis by the given angle $\arctan(x_i)$ and $\arctan(x_i^2)$, respectively. 
  The state is prepared with \emph{variational encoding}. The dashed box denotes one layer of a quantum subcircuit that is repeated twice. At the end of this circuit, two qubits are measured, and the $Z$ expectation values are calculated. The output from this circuit is a 2D vector. 
  }
\label{Fig:2D_VQC1}
\end{figure}
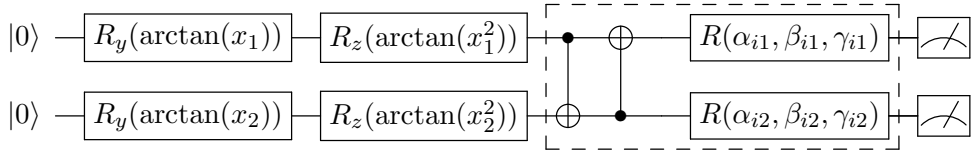
\begin{figure}[htbp]
\begin{center}
\begin{minipage}{10cm}
\Qcircuit @C=1em @R=1em {
\lstick{\ket{0}} & \gate{R_y(\arctan(x'_1))} & \gate{R_z(\arctan(x_1^{'2}))} & \ctrl{1}   & \targ      & \qw    & \gate{R(\alpha'_{i1}, \beta'_{i1}, \gamma'_{i1})} & \meter \qw \\
\lstick{\ket{0}} & \gate{R_y(\arctan(x'_2))} & \gate{R_z(\arctan(x_2^{'2}))} & \targ      & \ctrl{-1}  & \qw    & \gate{R(\alpha'_{i2}, \beta'_{i2}, \gamma'_{i2})} & \meter \qw 
\gategroup{1}{4}{2}{7}{.7em}{--}\qw 
}
\end{minipage}
\end{center}
\caption[Second VQC block for scikit 2D data classification]{{\bfseries Second quantum circuit block for 2D classification.}
  The parameters labeled $R_y(\arctan(x'_i))$ and $R_y(\arctan(x_i^{'2}))$ are for state preparation. $x'_1$ and $x'_2$ are the outputs of the first circuit block. The dashed box denotes one block of a quantum circuit that is repeated twice. 
  At the end of this circuit, two qubits are measured, and the $Z$ expectation values are calculated. The output from this circuit is a 2D vector. 
  In the context of cross-entropy loss, the outputs will be interpreted as the probability that the 2D point belongs to class one or two, respectively.}
\label{Fig:2D_VQC2}
\end{figure}
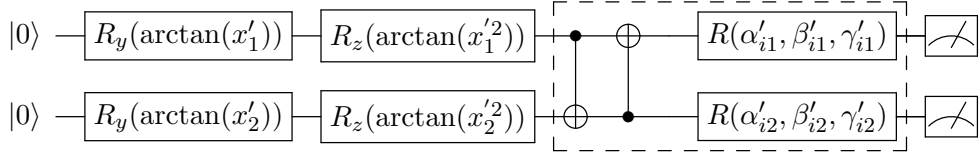

\begin{figure}[t]
\includegraphics[width=1\linewidth]{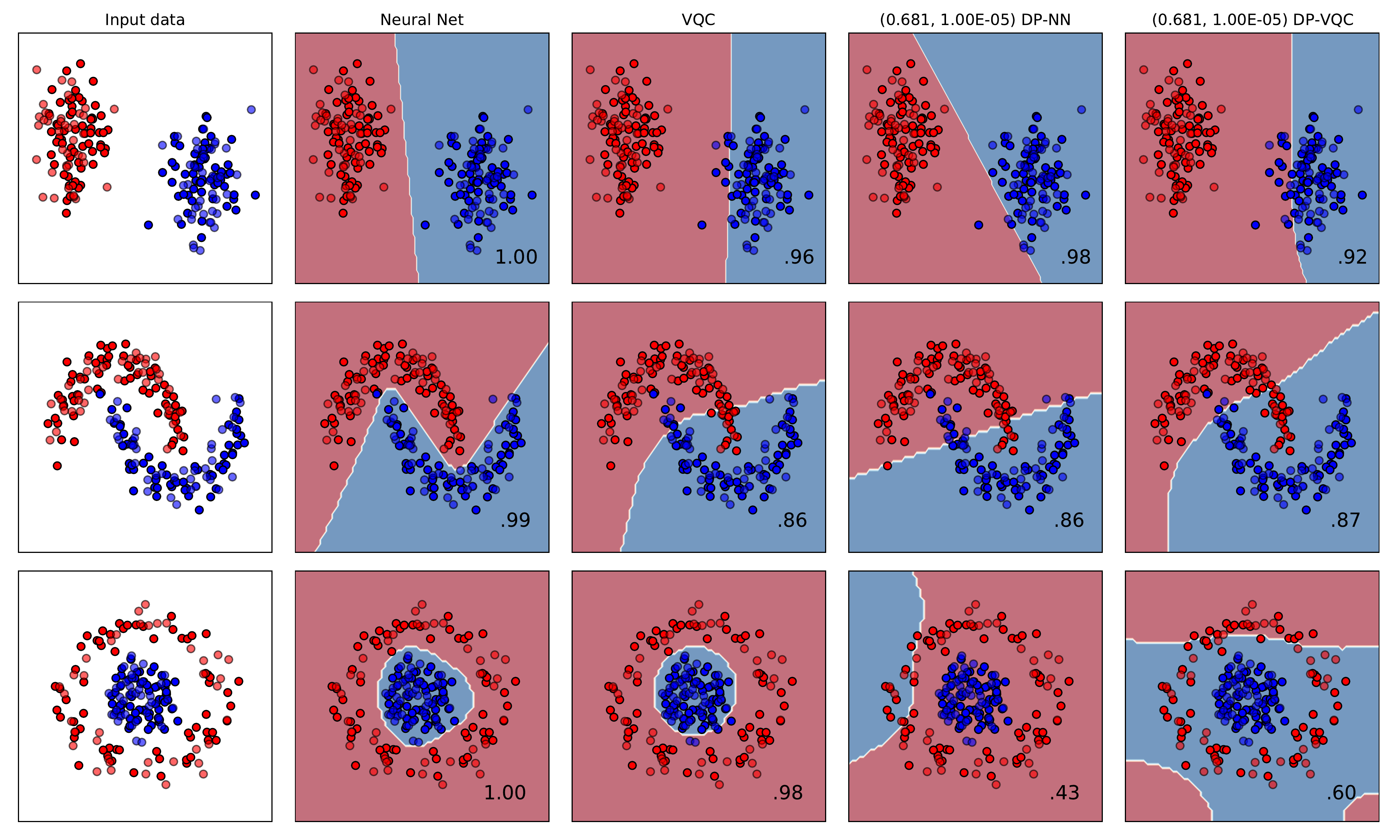}
\caption{Results from the 2D ML experiments. The first column shows three input sets. The subsequent columns show different models tasked with classifying and learning the decision boundary. The array of plots illustrates the decision boundaries formed by the different models. The solid points are those used in training, and the transparent ones are part of the testing set. Total accuracy after 30 epochs is displayed on the lower right of each plot.}
\label{fig:spatial}
\end{figure}

Three datasets of 2D classification from \texttt{scikit-learn} are considered. The generated datasets are divided into training, validating, and testing sets with 60\%, 20\%, and 20\% proportions, respectively. Different datasets are used because the decision boundary between the two classes is increasingly nonlinear and more difficult to classify. Thus, they make good benchmarks for DP training. The leftmost plots of \figureautorefname{\ref{fig:spatial}} display the input sets, which are named ``blobs,'' ``moons,'' and ``circles'' based on the shapes they form. The more transparent points are those not part of the training, but instead used for testing the model's accuracy.

As a baseline for the study, \figureautorefname{\ref{fig:nn_arch}} illustrates the classical neural network written with two classical layers. The classical classifier uses $\tanh$ as the activation function after each layer and softmax at the end of the calculation. The neural network has Xavier weight initialization. The linear layers sizes are such that the number of total trainable parameters in the quantum classifier is 66\%, while the number of trainable parameters in the classical classifier is $24$ for the VQC and $36$ for the neural network. 

The VQC to classify the 2D test set consists of two successive quantum subcircuits (Figures \ref{Fig:2D_VQC1} and \ref{Fig:2D_VQC2}). Each quantum subcircuit has two wires, while each unitary transform can be thought of as rotations on each qubit. Thus, each subcircuit is parameterized by $12$ Euler angles or parameters because there are two layers of transforms per subcircuit. The angles are initialized on a normal distribution with mean 0, standard deviation 1.0, and then scaled by 0.01.

Table \ref{table:spatial_eps} summarizes the key results from the 2D classification experiments. Three different levels of privacy have been investigated non-private, $(1.628, 10^{-5})$-DP, and $(0.681, 10^{-5}))$-DP on three different input sets blobs, moons, and circles. For most pairs of model architecture and input set, the differentially private result has a lower accuracy than the non-private one. The one exception is that the VQC classifies the moons more accurately with $(1.628, 10^{-5})$-DP than without privacy.

\begin{figure}[htbp]
\includegraphics[width=1\linewidth]{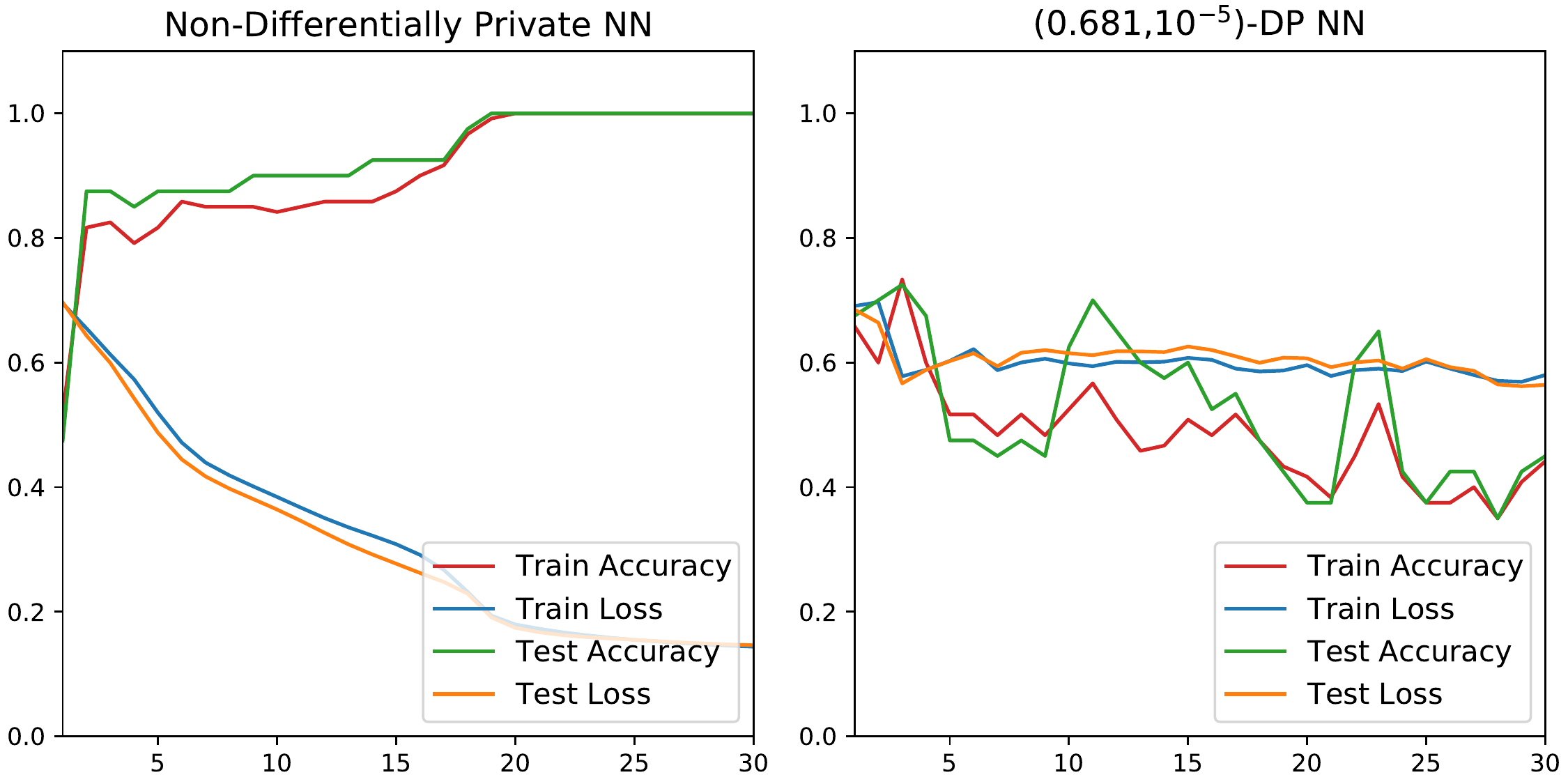} 
\caption{Results for ``circles'' classical classifier with 200 samples, a learning rate of 0.05, and RMSprop optimizer with and without DP.}
\label{fig:analysis_comparison_circles_c}
\end{figure}

\begin{figure}[htbp]
\includegraphics[width=1\linewidth]{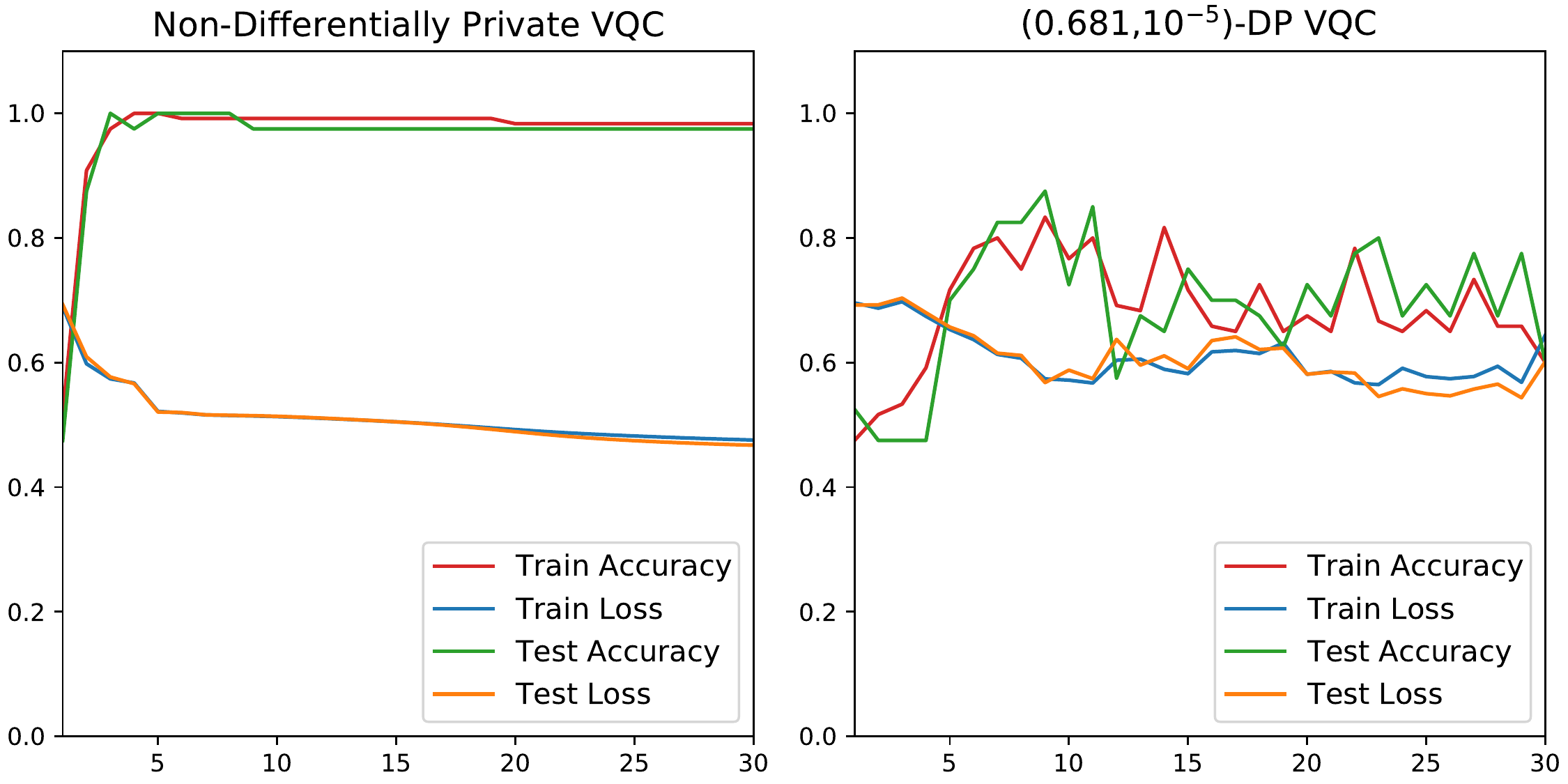} 
\caption{Results for ``circles'' variational quantum classifier with 200 samples, a learning rate of 0.05, and RMSprop optimizer with and without DP.}
\label{fig:analysis_comparison_circles_vqc}
\end{figure}

As detailed in Table \ref{table:spatial_eps}, the classical and quantum classifiers are almost equally successful for the blobs and moons sets. On the other hand, Figures \ref{fig:analysis_comparison_circles_vqc} and \ref{fig:analysis_comparison_circles_c} demonstrate that DP-VQC affords superior performance for the circles set as the quantum classifier's accuracy is between 13\% and 17\% higher than the DP-neural network. The last two columns of \figureautorefname{\ref{fig:spatial}} depict the decision boundary and accuracy of privacy-preserving $(0.681, 10^{-5})$-differentially private classical neural networks and VQCs. 

\begin{figure}[htbp]
\includegraphics[width=1\linewidth]{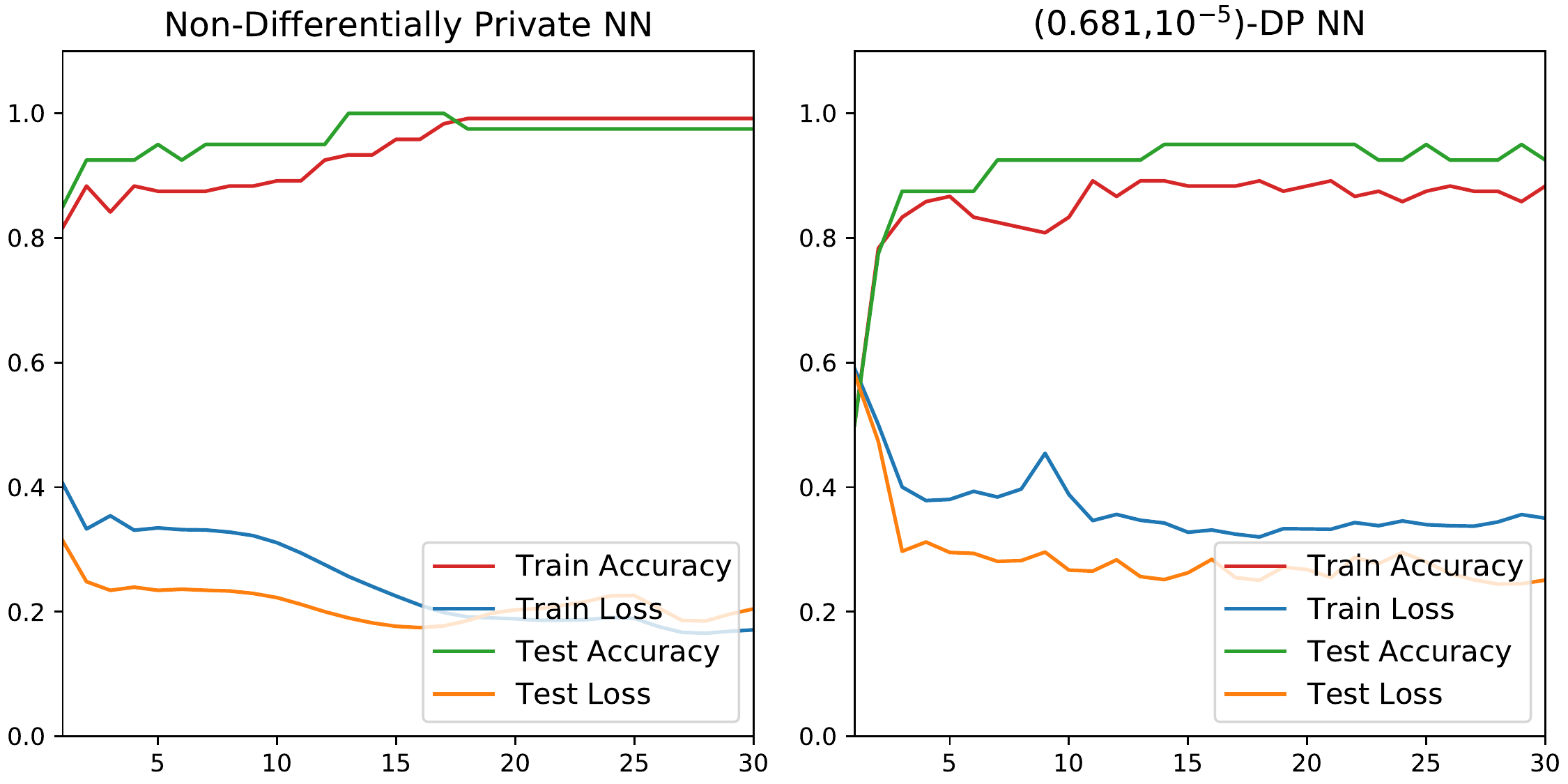} 
\caption{Results for ``moons'' classical classifier with 200 samples, a learning rate of 0.05, and RMSprop optimizer with and without DP.}
\label{fig:analysis_comparison_moons_c}
\end{figure}

The comparison of \figureautorefname{\ref{fig:analysis_comparison_moons_c}} and \figureautorefname{\ref{fig:analysis_comparison_circles_c}} demonstrates that the neural network's efficiency under DP training differs for different datasets. For the moons input set, the accuracy degradation from DP is somewhat significant at 10\%. Yet with the circles set, the accuracy decreases by ~40\%, and the final loss is double that of the non-private loss. Figures \ref{fig:analysis_comparison_circles_c} and \ref{fig:analysis_comparison_circles_vqc} illustrate that the VQC converges faster than the classical classifier, implying a potential quantum advantage over a classical neural network.

\begin{table}[ht]
\begin{center}
 \begin{tabular}{||c | c || c | c || c | c || c | c||} 
 \hline
 $\varepsilon$ & $\delta$ & NN-blobs & VQC-blobs & NN-moons & VQC-moons & NN-circles & VQC-circles \\ [0.5ex] 
 \hline\hline
 non-DP & n/a & 1.00 & 0.96 & 0.99 & 0.87 & 1.00 & 0.98 \\ 
 \hline
 1.628 & $10^{-5}$ & 0.98 & 0.92 & 0.91 & 0.88 & 0.47 & 0.60 \\
 \hline
 0.681 & $10^{-5}$ & 0.98 & 0.92 & 0.87 & 0.85 & 0.43 & 0.60 \\ [1ex] 
 \hline
\end{tabular}
\end{center}
\caption{Accuracies of differentially private neural networks and variational quantum classifiers after 30 epochs for 2D input sets: ``blobs,'' ``moons,'' and ``circles.'' The quantum classifier can achieve DP with more accuracy for the circles set. For the blobs and moons, the quantum and classical classifier has nearly the same accuracy under a given level of DP.}
\label{table:spatial_eps}
\end{table}

\subsection{MNIST Binary Classification}
\begin{figure}[htbp]
\begin{center}
\begin{minipage}{10cm}
\Qcircuit @C=1em @R=1em {
\lstick{\ket{0}} & \multigate{9}{U(\mathbf{x})}   & \ctrl{1}   & \qw       & \qw      & \qw      & \qw      & \qw      & \qw      & \qw      & \qw     & \qw      & \targ    &\gate{R(\alpha_1, \beta_1, \gamma_1)} & \meter \qw \\
\lstick{\ket{0}} & \ghost{U(\mathbf{x})}         & \targ      & \ctrl{1}  & \qw      & \qw      & \qw      & \qw      & \qw      & \qw      & \qw      & \qw      & \qw      &\gate{R(\alpha_2, \beta_2, \gamma_2)} & \meter \qw \\
\lstick{\ket{0}} & \ghost{U(\mathbf{x})}         & \qw        & \targ     & \ctrl{1} & \qw      & \qw      & \qw      & \qw      & \qw      & \qw      & \qw      & \qw      &\gate{R(\alpha_3, \beta_3, \gamma_3)} & \meter \qw \\
\lstick{\ket{0}} & \ghost{U(\mathbf{x})}         & \qw        & \qw       & \targ    & \ctrl{1} & \qw      & \qw      & \qw      & \qw      & \qw      & \qw      & \qw      &\gate{R(\alpha_4, \beta_4, \gamma_4)} & \meter \qw \\
\lstick{\ket{0}} & \ghost{U(\mathbf{x})}         & \qw        & \qw       & \qw      & \targ    & \ctrl{1} & \qw      & \qw      & \qw      & \qw      & \qw      & \qw      &\gate{R(\alpha_5, \beta_5, \gamma_5)} & \qw  \\
\lstick{\ket{0}} & \ghost{U(\mathbf{x})}         & \qw        & \qw       & \qw      & \qw      & \targ    & \ctrl{1} & \qw      & \qw      & \qw      & \qw      & \qw      &\gate{R(\alpha_6, \beta_6, \gamma_6)} & \qw  \\
\lstick{\ket{0}} & \ghost{U(\mathbf{x})}         & \qw        & \qw       & \qw      & \qw      & \qw      & \targ    & \ctrl{1} & \qw      & \qw      & \qw      & \qw      &\gate{R(\alpha_7, \beta_7, \gamma_7)} & \qw \\
\lstick{\ket{0}} & \ghost{U(\mathbf{x})}         & \qw        & \qw       & \qw      & \qw      & \qw      & \qw      & \targ    & \ctrl{1} & \qw      & \qw      & \qw      &\gate{R(\alpha_8, \beta_8, \gamma_8)} & \qw \\
\lstick{\ket{0}} & \ghost{U(\mathbf{x})}         & \qw        & \qw       & \qw      & \qw      & \qw      & \qw      & \qw      & \targ    & \ctrl{1} & \qw      & \qw      &\gate{R(\alpha_9, \beta_9, \gamma_9)} & \qw \\
\lstick{\ket{0}} & \ghost{U(\mathbf{x})}         & \qw        & \qw       & \qw      & \qw      & \qw      & \qw      & \qw      & \qw      & \targ    & \qw      & \ctrl{-9}&\gate{R(\alpha_{10}, \beta_{10}, \gamma_{10})} &  \qw \gategroup{1}{3}{10}{14}{.7em}{--}
}
\end{minipage}
\end{center}
\caption[First VQC block for MNIST classification]{{\bfseries First quantum circuit block for MNIST classification.}
 The first VQC block encodes the MNIST image. The $1024$-dimensional vector is encoded via amplitude encoding into a $\log(1024)$, i.e., $10$-qubit state. $U(\mathbf{x})$ denotes the quantum algorithm for amplitude encoding as explained in \cite{mottonen2005transformation, Schuld2018InformationEncoding}. $\alpha_{i}$, $\beta_{i}$, and $\gamma_{i}$ are the parameters to optimize. The dashed box denotes one block of a quantum circuit that is repeated eight times. Thus, there are $30 \times 8 = 240$ parameters to the circuit block. The dial to the far right represents that the circuit has four outputs. The expectation of $\sigma_z$ is measured on four qubits. The output becomes the input for the next circuit block.
 }
\label{Fig:MNIST_VQC1}
\end{figure}
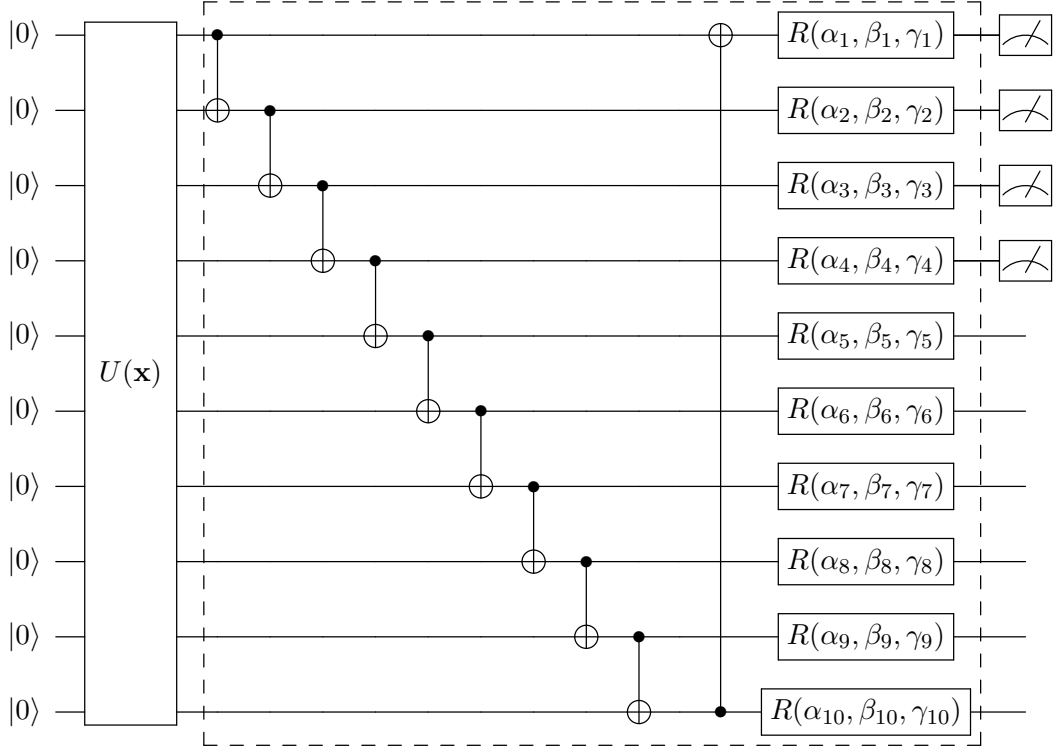
\begin{figure}[htbp]
\begin{center}
\begin{minipage}{10cm}
\Qcircuit @C=1em @R=1em {
\lstick{\ket{0}} & \gate{R_y(\arctan(x'_1))} &\gate{R_z(\arctan(x_1^{'2}))} & \ctrl{1}   & \qw    & \qw    & \targ     & \gate{R(\alpha_1, \beta_1, \gamma_1)} & \meter \qw \\
\lstick{\ket{0}} & \gate{R_y(\arctan(x'_2))} & \gate{R_z(\arctan(x_2{'2}))} & \targ      & \ctrl{1}  & \qw  & \qw   & \gate{R(\alpha_2, \beta_2, \gamma_2)} &  \meter \qw \\
\lstick{\ket{0}} & \gate{R_y(\arctan(x'_3))} & \gate{R_z(\arctan(x_3{'2}))} & \qw      & \targ     & \ctrl{1} & \qw   & \gate{R(\alpha_3, \beta_3, \gamma_3)} &  \qw \\
\lstick{\ket{0}} & \gate{R_y(\arctan(x'_4))} & \gate{R_z(\arctan(x_4{'2}))} & \qw      & \qw  & \targ     & \ctrl{-3}    & \gate{R(\alpha_4, \beta_4, \gamma_4)} &  \qw 
\gategroup{1}{4}{4}{8}{.7em}{--}\qw 
}
\end{minipage}
\end{center}
\caption[Second VQC block for MNIST classification]{{\bfseries Second quantum circuit block for MNIST classification.}
 The second subcircuit uses \emph{variational encoding} to encode the output from the first block to be the input for this subcircuit. $\alpha'_{i}$, $\beta'_{i}$, and $\gamma'_{i}$ are the parameters to optimize. The dashed box denotes one block of a quantum circuit that is repeated four times. There are $12 \times 4 = 48$ parameters to the circuit block. The dial to the far right represents that the circuit has four outputs, and the expectation of $\sigma_z$ is measured on two qubits. In the context of cross-entropy, the outputs will be interpreted as the probability that the image is of a `0' or a `1,' respectively.
 }
\label{Fig:MNIST_VQC2}
\end{figure}
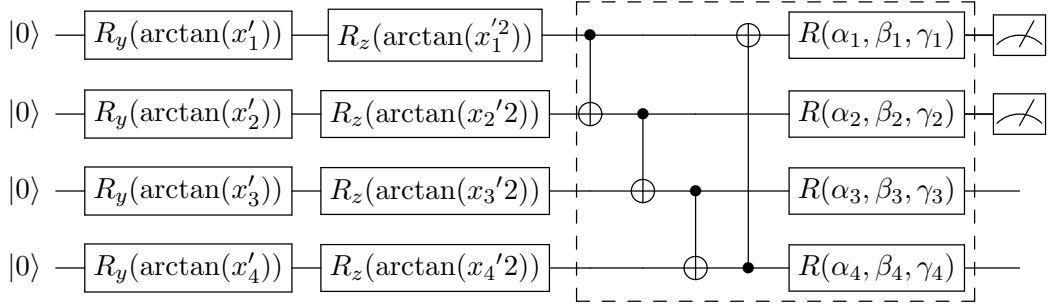

The MNIST classification task is prepared similarly to the 2D classification problem. Because of the computational complexity of simulating large quantum systems, the problem is reduced to a binary classification of distinguishing the handwritten digits of `0' and `1.' The digits are grayscale images with a total of 784 pixels. The variational quantum classifier uses \emph{amplitude loading} (described in \sectionautorefname{\ref{sec:AmplitudeEncoding}}) to compress the number of inputs to fit within 10 qubits. Therefore, the 784 inputs are padded with additional zeros to make the inputs 1024 dimensional. Next, \emph{amplitude loading} transforms the 1024 pixels into a $10$-qubit quantum state for operating the variational quantum classifier. 

The neural network uses the same padded 1024 pixels as an input. The hidden layer has only one node, and the output layer is two nodes. Hence, the classical model has 1029 parameters divided between the two weight matrices and biases. The design for this classical benchmark aims to limit the number of parameters for fair comparison to the quantum model.

\begin{figure}[htbp]
\includegraphics[width=1\linewidth]{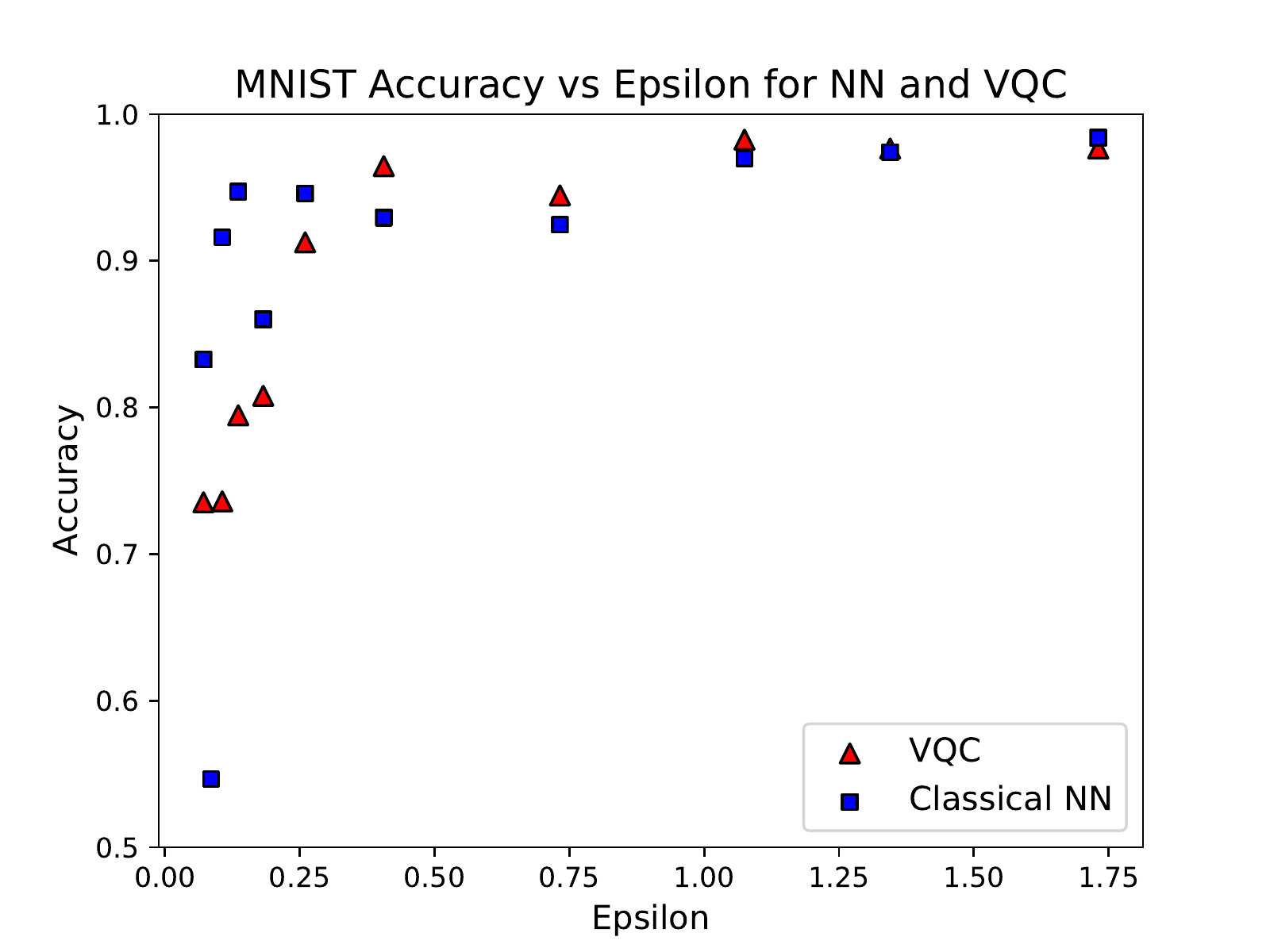}
\caption{Accuracy of DP classifiers after 30 epochs plotted against their $\varepsilon$ for a MNIST binary classification of the handwritten digits `0' and `1.' With only 28\% the number of parameters, the quantum classifier is at least as accurate as the classical classifier for all investigated DP levels.}
\label{fig:eps_plot}
\end{figure}

The quantum classifier has two quantum subcircuits. The first has 10 inputs, eight layers of unitary transforms, and four outputs (Figure \ref{Fig:MNIST_VQC1}). Each qubit has a tunable unitary transform per layer, so there are $8 \times 10 \times 3 = 240$ parameters in the first subcircuit. The second subcircuit has four inputs, two outputs, and four layers (Figure \ref{Fig:MNIST_VQC2}), so it has $4 \times 4 \times 3 = 48$ tunable parameters associated with the rotations of quantum bits. Consequently, the VQC has $288$ parameters. Importantly, this represents roughly only a quarter (27.99\% exactly) of the number of parameters associated with the analogous classical neural network used for the same classification task.
The MNIST results are summarized in Table \ref{table:eps}. Multiple levels of privacy are created by iterating the noise multiplier from $1.0$ to $5.0$. The privacy budget for such noise is between $1.73$ to $0.07$, respectively. \figureautorefname{\ref{fig:eps_plot}} and Table \ref{table:eps} exemplify that the accuracies of both neural networks and VQCs decrease as $\varepsilon$ decreases. This emphasizes the trade-off between utility and privacy in differentially private algorithms.

\begin{table}[htbp]
 \begin{tabular}{||c | c | c | c||} 
 \hline
 $\epsilon$ & $\delta$ & Classical NN & VQC \\ [0.5ex] 
 \hline\hline
 1.73071508 & $10^{-5}$ & 0.984 & 0.97666667 \\ 
 \hline
 1.3448161 & $10^{-5}$ & 0.974 & 0.97666667 \\
 \hline
 1.07469683 & $10^{-5}$ & 0.97 & 0.98266667 \\
 \hline
 0.73250501 & $10^{-5}$ & 0.92466667 & 0.94466667 \\
 \hline
 0.40585425 & $10^{-5}$ &  0.92933333 & 0.96466667 \\ 
 \hline
 0.25998742 & $10^{-5}$ & 0.946 & 0.91266667 \\ 
 \hline
 0.18230998 & $10^{-5}$ &  0.86 & 0.808 \\
 \hline
 0.13604452 & $10^{-5}$ &  0.94733333 & 0.79466667  \\
 \hline
 0.10626109 & $10^{-5}$ & 0.916 & 0.736  \\
 \hline
 \hline
 0.07149769 & $10^{-5}$ & 0.83266667 & 0.73533333 \\ [1ex] 
 \hline
\end{tabular}
\caption{Results from binary MNIST classification. Accuracies of differentially private neural networks and variational quantum classifiers after 30 epochs. 
The private quantum classifier is more accurate and successful for $\varepsilon$s between 0.41 and 1.34.}
\label{table:eps}
\end{table}

\begin{table}[htbp]
 \begin{tabular}{||c || c | c ||} 
 \hline
 model & DP-NN & DP-VQC \\ [0.5ex] 
 \hline\hline
 \# of parameters & 1029 & 288 \\ 
 \hline
 Final accuracy & 0.929 & 0.965 \\
 \hline
\end{tabular}
\caption{Results from binary MNIST classification for a $(0.406, 10^{-5})$-differentially private neural network (DP-NN) and VQC (DP-VQC).}
\label{table:mnistFinalAccs}
\end{table}

\section{\label{sec:Discussion}Discussion}
%


\subsection{Potential Applications of Private QML}
Differentially private data are becoming more critical because larger models have been shown to memorize more data, e.g., language models \cite{carlini2020extracting}. One of the latest state-of-the-art language models, GPT-2, has 1.5 billion parameters and was found to memorize 18 times more information when compared to a 124 million parameters language model. The aforementioned study demonstrates that training data extraction attacks are practical. This necessitates an implementation of a privacy-preserving algorithm, i.e., DP, to curtail memorization and data extraction attacks.

This study has presented the implementation and successful roof-of-concept application of DP to QML models. The application can be extended to a myriad of applications that require privacy-preserving learning and the power advantage stemming from QML. One potential application is facial recognition. These models must train on thousands faces, whose identities are not protected. Therefore, this area would intrinsically benefit from DP, and QML could create even more accurate predictions \cite{easom2020towards}. QCNN would be another logical application of a private QML algorithm as QCNNs already are being investigated with the MNIST and other benchmarks \cite{liu2019hybrid, chen2021qgcnn, kerenidis2019quantum}. Our results show that the private VQC distinguishes between the `0' and `1' digits with an accuracy exceeding 90\%. As such, it is expected that a privacy-preserving framework would benefit these application scenarios, including QCNN.

With recent QML developments impacting a spectrum of applications, such as speech recognition \cite{yang2020decentralizing}, quantum recurrent neural network (QRNN) and quantum LSTM for sequential learning \cite{chen2020quantum, bausch2020recurrent, takaki2020learning}, and even certain emerging applications in medical imaging \cite{houssein2021hybrid}, we expect the framework described by this work would be of benefit to these new scenarios as well.

An important point to consider is the limitation of extending these results to real-world quantum devices. High-dimensional input, such as MNIST, takes an extremely long time to run on a cloud-based quantum computer. However, it is possible to run. One limitation of this study is that the VQCs are simulated with noise-free quantum computers. A future study could investigate the results of running privacy-preserving quantum optimization on a noisy simulator or cloud-based quantum computer.

\subsection{Success of Differentially Private QML}
This research is seeking to show that a differentially private variational quantum classifier can be trained to identify the decision boundary between two classes. \figureautorefname{\ref{fig:spatial}} shows that the given hyperparameters achieve nearly perfect classification. After 30 epochs, both the quantum and classical classifiers achieve accuracies greater than 95\% for data organized into blobs and concentric circles. On the other hand, the classical network achieves 99\% accuracy for the moons classification, while the moons dataset proved to be the most difficult input for the quantum classifier to classify, achieving merely 86\% accuracy. It may be conjectured that the VQC had difficulties in learning the highly convex decision boundary necessary for the moons input set. In spite of that, the VQC generally trains just as well as a classical neural network with only 66\% of the total parameters.

While DP training usually causes models to fail to capture the long tail of a data distribution, the DP-QML training is just as successful as the non-private algorithm for the blobs and moons datasets, where only a modest accuracy penalty occurs. While the accuracy of the private training for the circles classification is much lower than its non-private counterpart, the DP-VQC still is much more successful at the task than the classical differentially private neural network. Our study demonstrates that a quantum advantage can offset the usual compromise between privacy and accuracy as seen in other DP applications \cite{deepLearningDP, disparateImpact}.

The MNIST binary classification problem creates an even more compelling case for the QML algorithm being advantageous compared to a classical ML algorithm. \figureautorefname{\ref{table:mnistFinalAccs}} demonstrates that a privacy-preserving variational quantum classifier can learn to distinguish between the handwritten digits `0' and `1' from the MNIST dataset to an accuracy of nearly 100\%. The same figure shows that a classical neural network also can accomplish the task. The quantum advantage arises because the quantum network has only 288 parameters compared to the 1029 parameters characterizing the classical neural network. Furthermore, the differentially private VQC attains better accuracy than the classical neural network for $\varepsilon$s between 0.4 and 1.4 (shown in Table \ref{table:eps}). This range of $\varepsilon$ is sufficient, where differentially private techniques attain good privacy as defined in \cite{deepLearningDP}. 

This work mainly focuses on the numerical demonstration of potential quantum advantages, leaving the theoretical investigation for future work.

\section{\label{sec:Conclusion}Conclusion}
Overall, the QML algorithm attains the same accuracy in the MNIST classification task as the classical ML algorithm with only 28\% of the number of parameters, making it more efficient than an ML algorithm. In this work, a QML algorithm in a differentially private framework is developed, and the quantum advantage is maintained when the ML algorithm is improved to preserve privacy. This research also shows that VQCs maintain their quantum advantage under DP in the classification of the handwritten digits `0' and `1' and 2D nonlinear classifications with careful selection of hyperparameters.

This novel framework combines differentially private optimization with QML. Including DP in the algorithm ensures privacy-preserving learning. We also demonstrate a capacity for high-fidelity privacy and high accuracy in variational quantum classifiers with two different benchmarks. Notably, we show the superior performance in terms of convergence of differentially private QML over classical DP-ML. These results indicate the potential benefits quantum computing will bring to privacy-preserving data analytics.

\begin{acknowledgments}
This work was supported by the U.S.\ Department of Energy (DOE), Office of Science (SC), Advanced Scientific Computing Research program under award DE-SC-0012704 and Brookhaven National Laboratory's Laboratory Directed Research and Development Program (\#20-024).
This project also was supported in part by the DOE-SC's Office of Workforce Development for Teachers and Scientists (WDTS) under the Science Undergraduate Laboratory Internships Program (SULI). 
\end{acknowledgments}

\appendix
\section{Software Packages}
\subsection{\label{app:PyvacyLossCalc}Privacy Loss Calculator}
The package used in this research, \texttt{PyVacy}, implements a privacy loss calculation based on the TensorFlow privacy calculator \cite{pyvacy}. First, the Renyi differential privacy (RDP) epsilon and order are calculated. Then, the RDP loss is determined. Usually, this is more feasible because a composition property holds for $(\alpha, \varepsilon)$ RDP (shown in Proposition 1 of \cite{renyi}). Thus, the Renyi divergences can be exactly added together given the same order. At the end of the calculation, the $(\alpha, \varepsilon)$ RDP is converted into the $(\varepsilon, \delta)$-differential privacy loss.
$(\alpha, \varepsilon)$-RDP is defined by the Renyi divergence~\cite{renyi}:
\begin{equation}
    \mathrm{D}_\alpha \equiv \frac{1}{\alpha-1}\ln E_{x \sim Q} \left(  \frac{P(x)}{Q(x)} \right).
\end{equation} 

The weights for the classical neural networks are initialized uniformly from minus to plus $\frac{1}{\sqrt{7}}$ and $\frac{1}{\sqrt{2}}$ for the first and second matrices, respectively. For the variational quantum classifiers, the angle parameters of the unitary transforms are sampled from a normalized distribution scaled by 0.01.

\subsection{\label{app:PyvacyML}Differentially Private Machine Learning}
A micro-batch of size $m$ and mini-batch of size $n$ are defined, where $n/m$ micro-batches make up a mini-batch. The loss and its gradient are calculated for each micro-batch, and a total norm of gradients is calculated. The micro-batch gradients are scaled and summed together with additional noise. This accumulated gradient now is the effective gradient for the mini-batch. This creates DP by limiting the effect each training point has on the batch gradient.

The Python package \texttt{PyVacy} is used to add DP to the RMSprop optimizer employed in this study. The DP optimizer crucially overrides the step() method and defines a micro-batch\_step(). Normally the loss and its gradients are calculated for each mini-batch. In \texttt{PyVacy}, each mini-batch is split into micro-batches. In each micro-batch, the loss and loss gradient are calculated. Then, the micro-batch step function is called, clipping the gradients. After all the micro-batches, the step function is called to add Gaussian noise and update the parameters according to these altered gradients \cite{pyvacy}.

The mini-batch step takes the parameter gradients for each micro-batch and calculates the effective mini-batch gradients. First, the total norm of the parameters gradients, $N$, is calculated. Then, the scaled micro-batch gradients are added to a new parameter, called the \textit{accumulated gradient}. The gradients are scaled by a coefficient that scales down the gradients to have a total norm equal to the norm cutoff, $S$, or, if necessary, $c = \min (\frac{S}{N + 1e-6}, 1)$. The accumulated gradients add the micro-batch gradients together to create a new effective gradient for the mini-batch. 

This effective gradient is scaled so that the loss gradients, calculated from a given micro-batch, do not have too large norms. This creates DP by limiting the effect each training point has on the batch gradient \cite{deepLearningDP}. In the overridden step method, the accumulated gradients have Gaussian noise added. The Gaussian noise is proportional to the norm cutoff and noise multiplier, $S$ and $z$, respectively. The accumulated gradient then is scaled by the ratio of micro-batch to mini-batch sizes, and the micro-batch size usually is set to be 1. This has the effect of using the accumulated gradients in place of the original parameter gradients in the step update rule. 

\bibliographystyle{ieeetr}
\bibliography{DP-QVC,bib/nisq,bib/qml_examples,bib/tools,bib/vqc,bib/ml,bib/qc}

\end{document}